\documentclass[onecolumn,peerreview]{IEEEtran}
\ifCLASSINFOpdf
\else
    \usepackage[dvips]{graphicx}
\fi
\usepackage{amsmath,amssymb}
\newcommand{\bysame}{%
    \leavevmode\hbox to 3em{\hrulefill}\,}

\begin{document}
%
\title{Characteristic Matrices and Trellis Reduction for Tail-Biting Convolutional Codes}
%
%
%

\author{Masato~Tajima,~\IEEEmembership{Senior~Member,~IEEE}
\thanks{M. Tajima was with the graduate School of
Science and Engineering, University of Toyama, 3190 Gofuku,
Toyama 930-8555, Japan (e-mail: masatotjm@kind.ocn.ne.jp).}
\thanks{Manuscript received April 19, 2005; revised August 26, 2015. This paper was presented in part at the IEICE Technical Committee IT Conference in March 2017.}}

%
%

\markboth{Journal of \LaTeX\ Class Files,~Vol.~14, No.~8, August~2015}%
{Tajima \MakeLowercase{\textit{et al.}}: Characteristic Matrices and Trellis Reduction for Tail-Biting Convolutional Codes}
%



\maketitle

\begin{abstract}
Basic properties of a characteristic matrix for a tail-biting convolutional code are investigated. A tail-biting convolutional code can be regarded as a linear block code. Since the corresponding scalar generator matrix $G^{tb}$ has a kind of cyclic structure, an associated characteristic matrix also has a cyclic structure, from which basic properties of a characteristic matrix are obtained. Next, using the derived results, we discuss the possibility of trellis reduction for a given tail-biting convolutional code. There are cases where we can find a scalar generator matrix $G'$ equivalent to $G^{tb}$ based on a characteristic matrix. In this case, if the polynomial generator matrix corresponding to $G'$ has been reduced, or can be reduced by using appropriate transformations, then trellis reduction for the original tail-biting convolutional code is realized. In many cases, the polynomial generator matrix corresponding to $G'$ has a monomial factor in some column and is reduced by dividing the column by the factor. Note that this transformation corresponds to cyclically shifting the associated code subsequence (a tail-biting path is regarded as a code sequence) to the left. Thus if we allow partial cyclic shifts of a tail-biting path, then trellis reduction is accomplished.
\end{abstract}

\begin{IEEEkeywords}
Tail-biting convolutional codes, tail-biting trellis, characteristic matrix, cyclic shift, trellis reduction.
\end{IEEEkeywords}

%
\IEEEpeerreviewmaketitle

\section{Introduction}
%
%
%
%
From the 1980s to 1990s, trellis representations of linear block codes were studied with a great interest~\cite{bahl 74,forn 88,forn 94,ks 95,mc 96,mc 962,mud 88}. Subsequently, tail-biting trellises of linear block codes have received much attention. Given a linear block code, there exists a unique minimal conventional trellis. This trellis simultaneously minimizes all measures of trellis complexity. However, tail-biting trellises do not have such a property. That is, minimality of tail-biting trellises depends on the measure being used~\cite{koe 03}. In general, the complexity of a tail-biting trellis may be much lower than that of the minimal conventional trellis. There have been many contributions to the subject, including~\cite{cal 99,con 15,glu 111,glu 112,glu 13,koe 03,lin 00,nori 06,sha 00,wea 12}. The works~\cite{cal 99,koe 03} had a strong influence on the subsequent studies. A remarkable progress has been made by Koetter and Vardy in their paper~\cite{koe 03}. They showed that for a $k$-dimensional linear block code of length $n$ with full support, there exists a list of $n$ characteristic generators (i.e., a characteristic matrix~\cite{koe 03}) from which all minimal tail-biting minimal trellises can be obtained. A different method of producing tail-biting trellises was proposed by Nori and Shankar~\cite{nori 06}. They used the {\it Bahl-Cocke-Jelinek-Raviv} (BCJR) construction~\cite{bahl 74}. These works were further investigated by Gluesing-Luerssen and Weaver~\cite{glu 111,glu 112}. In particular, noting that a characteristic matrix for a given code is not necessarily unique, they have refined and generalized the previous works. More recent works~\cite{con 15,glu 13} provide further research on the subject.
\par
On the other hand, {\it tail-biting convolutional codes} were proposed by Ma and Wolf in 1986~\cite{ma 86} (tail-biting representations of block codes were introduced by Solomon and van Tilborg~\cite{solo 79}). {\it Tail-biting} (abbreviated TB) is a technique by which a convolutional code can be used to construct a block code without any loss of rate. In connection with the subject, there have been also many works, including~\cite{and 02,ma 86,shao 03,sta 99,taji 122,wei 01}. Since a TB convolutional code is identified with a linear block code, the results on TB trellises for linear block codes can be used. In particular, we can think of a characteristic matrix of a given TB convolutional code. In this paper, we first investigate a characteristic matrix for a TB convolutional code. And then, based on the derived results, we discuss the possibility of trellis reduction for a given TB convolutional code. An outline of the rest of the paper is as follows:
\par
In Section II, we review the basic notions needed for this paper.
\par
In Section III, we investigate the basic properties of a characteristic matrix for a TB convolutional code. When a TB convolutional code with generator matrix $G(D)$ is regarded as a linear block code $C$, a (scalar) generator matrix (denoted by $G^{tb}$) for $C$ is constructed using the coefficients which appear in the polynomial expansion of $G(D)$. We see that $G^{tb}$ has a kind of cyclic structure. Then it is shown that the (characteristic) span list associated with a characteristic matrix for $C$ consists of some basic spans and their right cyclic shifts, from which basic properties of a characteristic matrix are derived.
\par
In Section IV, we deal with transformations of $G(D)$ and discuss the relationship between these transformations and the corresponding scalar generator matrices $G^{tb}\mbox{'}s$. We see that dividing a column of $G(D)$ by a monomial factor corresponds to cyclically shifting a column subsequence of $G^{tb}$ to the left, whereas multiplying a column of $G(D)$ by a monomial corresponds to cyclically shifting a column subsequence of $G^{tb}$ to the right. These properties are essentially used for trellis reduction to be discussed in Section V.
\par
In Section V, we discuss the possibility of trellis reduction for a given TB convolutional code (we identify the code with an $(n, k)$ block code $C$). As is stated above, we can think of a characteristic matrix for $C$. Consider the case where some $k$ characteristic generators, which consist of some basic generators and their right cyclic shifts, can generate the same code $C$. We see that these characteristic generators form a (scalar) generator matrix associated with a (polynomial) generator matrix of another convolutional code. In this case, if the constraint length of the obtained generator matrix is smaller than that of the original one, then trellis reduction is realized. Even if this kind of reduction is not possible, there are cases where a newly obtained generator matrix contains a monomial factor in some column. Then there is a possibility that the generator matrix is reduced by sweeping the monomial factor out of the column. Note that this operation corresponds to cyclically shifting the corresponding code subsequence to the left. In this way, trellis reduction can be accomplished. We also present a trellis reduction method for high rate codes which uses a reciprocal dual encoder. We remark that the (trellis) section length is an important parameter and the proposed method is restricted to TB convolutional codes with short to moderate section length. We give an upper bound for the section length by evaluating the span lengths of characteristic generators.
\par
Finally, conclusions are provided in Section VI.

\section{Preliminaries}
We begin with the basic notions needed in this paper, where the underlying field is assumed to be $F=\mbox{GF}(2)$. Let $C$ be an $(n, k)$ linear block code, where the set of indices for a codeword in $C$ is denoted by $I\stackrel{\triangle}{=}\{0, 1, \cdots, n-1\}$. Then a codeword $x \in C$ is expressed as $x=(x_0, x_1, \cdots, x_{n-1})$. $I$ is also regarded as the time axis for TB trellises for $C$. Since TB trellises for $C$ are considered in this paper, it is convenient to identify $I$ with $Z_n$, the ring of integers modulo $n$. Hence, when dealing with TB trellises, all index arithmetic will be implicitly performed modulo $n$~\cite{koe 03}.
\par
The notion of {\it span} is fundamental in trellis theory. Given a codeword $x \in C$, a span of $x$, denoted by $[x]$, is a semiopen interval $(a, b] \in I$ such that the corresponding closed interval $[a, b]$ contains all the nonzero positions of $x$~\cite{koe 03}. Due to the cyclic structure of the time axis $I$, we adopt the following interpretation of intervals~\cite{glu 111,glu 112,koe 03}. For $a, b \in I$, we define
\begin{equation}
[a, b]\stackrel{\triangle}{=}\left\{
\begin{array}{ll}
\{a, a+1, \cdots, b\}, & \quad \mbox{if $a \leq b$} \\
\{a, a+1, \cdots, n-1, 0, 1, \cdots, b\}, & \quad \mbox{if $a > b$}
\end{array}
\right.
\end{equation}
and $(a, b]\stackrel{\triangle}{=}[a, b]\backslash \{a\}$. We call the intervals $(a, b]$ and $[a, b]$ {\it conventional} if $a \leq b$ and {\it circular} otherwise.
\par
In connection with the construction of minimal TB trellises for $C$, Koetter and Vardy~\cite{koe 03} introduced the notion of {\it characteristic generator} for $C$. Denote by $\sigma_j(\cdot)$ a cyclic shift to the left by $j$ positions~\cite{koe 03}. Similarly, denote by $\rho_j(\cdot)$ a cyclic shift to the right by $j$ positions. Let $X_j^*$ be a basis in minimal-span form~\cite{mc 96} for the code $C_j\stackrel{\triangle}{=}\sigma_j(C)$. A characteristic generator for $C$ is a pair consisting of a codeword $x=(x_0, x_1, \cdots, x_{n-1})\in C$ and a span $[x]=(a, b]$ such that $x_a, x_b$ are nonzero. The set of all the characteristic generators for $C$ is given by
\begin{equation}
X\stackrel{\triangle}{=}X_0^* \cup \rho_1(X_1^*)\cup \cdots \cup \rho_{n-1}(X_{n-1}^*) .
\end{equation}
Here we have an understanding that if $x^* \in X_j^*$, then $[\rho_j(x^*)]=(a+j, b+j]$, where $[x^*]=(a, b]$.
\par
Assume that $C$ has a full support. Then a {\it characteristic matrix} for $C$ is the $n \times n$ matrix having the elements of $X$ as its rows. The above definition implies that when we refer to a characteristic matrix, the associated spans are taken into account. Here note that a basis in minimal-span form is not necessarily unique. Hence, $X$ may not be uniquely determined. On the other hand, the set of spans (denoted by $T$) accompanied by $X$ is, up to ordering, uniquely determined by the code $C$~\cite{glu 111,glu 112,koe 03}. $T$ is called the {\it characteristic span list} of $C$ (an element$\in T$ is called a characteristic span of $C$)~\cite{glu 111,glu 112}. In order to clarify this fact, Gluesing-Luerssen and Weaver introduced the notion of {\it characteristic pair} $(X, T)$ of $C$~\cite[Definition III.8]{glu 111}, where $X$ is a generating set of $C$ and $T$ represents the associated spans. In this paper, we basically follow the definition of Gluesing-Luerssen and Weaver, but in order to emphasize the fact that a characteristic matrix inherently assumes the associated spans, we leave the term {\it characteristic matrix} for the definition. Thus we define as follows (cf.~\cite[Definition III.8]{glu 111}).
\newtheorem{df}{Definition}[section]
\begin{df}
Let $C$ be an $(n, k)$ linear block code with support $I$. A characteristic matrix for $C$ with (characteristic) span list $T$ is defined to be a pair $(X, T)$, where
\begin{equation}
X=\left(
\begin{array}{c}
x_1  \\
x_2  \\
\cdots \\
x_n
\end{array}
\right)
\end{equation}
\begin{equation}
T=\{(a_l, b_l]:~l=1, 2, \cdots, n\}
\end{equation}
have the properties:
\begin{itemize}
\item[1)] $\{x_1, \cdots, x_n\}$ generates $C$.
\item[2)] $(a_l, b_l]$ is a span of $x_l,~l=1, \cdots, n$.
\item[3)] $a_1, \cdots, a_n$ are distinct and $b_1, \cdots, b_n$ are distinct.
\item[4)] For all $j \in I$, there exist exactly $n-k$ row indices, $l_1, \cdots, l_{n-k}$, such that $j \in (a_{l_i}, b_{l_i}]$ for $i=1, \cdots, n-k$.
\end{itemize}
\end{df}
\par
{\it Remark:} Property 3) is derived from~\cite[Lemma 5.7]{koe 03} and the related remarks. Also, Property 4) is derived from the proof of~\cite[Theorem 5.10]{koe 03}.
\par
In the following, when there is no danger of confusion, we shall use the terms {\it characteristic matrix $X$} and {\it characteristic matrix $X$ with span list $T$} interchangeably.

\section{Characteristic Matrices for a Tail-Biting Convolutional Code}
Let $G(D)$ be a polynomial generator matrix of size $k_0 \times n_0$. Denote by $H(D)$ a corresponding polynomial check matrix. Both $G(D)$ and $H(D)$ are assumed to be {\it canonical}~\cite{mc 962}. Consider a standard trellis of $N$ sections for a convolutional code defined by $G(D)$. Here $\max(L, M)+1 \leq N$ is assumed, where $L$ and $M$ are the memory lengths of $G(D)$ and $H(D)$, respectively. The TB condition is a restriction that the encoder starts and ends in the same state. That is, only those paths in the trellis that start and end in the same state are admissible. We call such paths {\it TB paths}. Let $C$ be the set of all TB paths. In the following, we call $C$ a TB convolutional code of section length $N$ defined by $G(D)$ (cf. Fig.1 in Section V). When there is no danger of confusion, we will omit the phrase {\it of section length $N$}. $C$ can be regarded as a linear block code $B^{tb}$ of length $n=n_0N$. To simplify the notations, $B^{tb}$ is identified with $C$ and is denoted simply by $C$. Let
\begin{equation}
G(D)=G_0+G_1D+\cdots +G_LD^L
\end{equation}
be the polynomial expansion of $G(D)$, where $G_i~(0 \leq i \leq L)$ are $k_0 \times n_0$ matrices. Then the scalar generator matrix for $C~(=B^{tb})$ is given by
\begin{equation}
\arraycolsep=1pt
G_N^{tb}\stackrel{\triangle}{=}\left(
\arraycolsep=1pt
\begin{array}{cccccccc}
G_0 & G_1 & \scriptstyle{\ldots} & G_{L-1} & G_L &  &  &  \\
 & G_0 & \scriptstyle{\ldots} & \scriptstyle{\ldots} & G_{L-1} & G_L &  &  \\
 &  & \scriptstyle{\ldots} & \scriptstyle{\ldots} & \scriptstyle{\ldots} & \scriptstyle{\ldots} & \scriptstyle{\ldots} &  \\
 &  &  & G_0 & G_1 & \scriptstyle{\ldots} & \scriptstyle{\ldots} & G_L \\
G_L &  &  &  & G_0 & G_1 & \scriptstyle{\ldots} & G_{L-1} \\
G_{L-1} & G_L &  &  &  & G_0 & \scriptstyle{\ldots} & \scriptstyle{\ldots} \\
\scriptstyle{\ldots} & \scriptstyle{\ldots} & \scriptstyle{\ldots} &  &  &  & \scriptstyle{\ldots} & G_1 \\
G_1 & G_2 & \scriptstyle{\ldots} & G_L &  &  &  & G_0
\end{array}
\right)
\end{equation}
with size $k \times n=k_0N \times n_0N$~\cite{joha 99}. Hence, we can say that a TB convolutional code $C$ is generated by $G_N^{tb}$. In the following, we call $G_N^{tb}$ the {\it tail-biting generator matrix} (abbreviated TBGM) associated with a TB convolutional code $C$ defined by $G(D)$, or simply the TBGM associated with $G(D)$.

\subsection{Computation of Characteristic Matrices}
Koetter and Vardy~\cite{koe 03} have given an algorithm which can compute a characteristic matrix for a linear block code. Consider a TB convolutional code $C$ generated by $G_N^{tb}$. Note that $\sigma_{n_0}(G_N^{tb})$ is equivalent to $G_N^{tb}$. That is, $G_N^{tb}$ has a periodic structure of period $n_0$. Using this property, a characteristic matrix $X$ for $C$ can be computed efficiently.
\par
Let $C_j\stackrel{\triangle}{=}\sigma_j(C)$. $C_j$ is the code generated by $\sigma_j(G_N^{tb})$. Let $X_j^*$ be a basis in minimal-span form for the code $C_j$. Then a characteristic matrix $X$ for $C$ is defined as follows~\cite{koe 03}:
\begin{eqnarray}
X &\stackrel{\triangle}{=}& X_0^*\cup \rho_1(X_1^*)\cup \cdots \cup \rho_{n_0-1}(X_{n_0-1}^*) \nonumber \\
&& \cup \rho_{n_0}(X_{n_0}^*)\cup \rho_{n_0+1}(X_{n_0+1}^*)\cup \cdots \cup \rho_{2n_0-1}(X_{2n_0-1}^*) \nonumber \\
&& \cdots \nonumber \\
&& \cup \rho_{(N-1)n_0}(X_{(N-1)n_0}^*)\cup \rho_{(N-1)n_0+1}(X_{(N-1)n_0+1}^*)\cup \cdots \cup \rho_{Nn_0-1}(X_{Nn_0-1}^*) .
\end{eqnarray}
Since $\sigma_{n_0}(G_N^{tb})$ is equivalent to $G_N^{tb}$, we have
\begin{eqnarray}
\rho_{n_0}(X_{n_0}^*) &=& \rho_{n_0}(X_0^*) \nonumber \\
\rho_{n_0+1}(X_{n_0+1}^*) &=& \rho_{n_0}(\rho_1(X_1^*))=\rho_{n_0}(\tilde X_1^*) \nonumber \\
&& \cdots \nonumber \\
\rho_{2n_0-1}(X_{2n_0-1}^*) &=& \rho_{n_0}(\rho_{n_0-1}(X_{n_0-1}^*))=\rho_{n_0}(\tilde X_{n_0-1}^*) , \nonumber
\end{eqnarray}
where
\begin{equation}
\left\{
\begin{array}{l}
\tilde X_1^*\stackrel{\triangle}{=}\rho_1(X_1^*) \\
\tilde X_2^*\stackrel{\triangle}{=}\rho_2(X_2^*) \\
\cdots \\
\tilde X_{n_0-1}^*\stackrel{\triangle}{=}\rho_{n_0-1}(X_{n_0-1}^*) .
\end{array} \right. 
\end{equation}
Similarly, we have
\begin{eqnarray}
\rho_{2n_0}(X_{2n_0}^*) &=& \rho_{2n_0}(X_0^*) \nonumber \\
\rho_{2n_0+1}(X_{2n_0+1}^*) &=& \rho_{2n_0}(\rho_1(X_1^*))=\rho_{2n_0}(\tilde X_1^*) \nonumber \\
&\cdots& \nonumber \\
\rho_{3n_0-1}(X_{3n_0-1}^*) &=& \rho_{2n_0}(\rho_{n_0-1}(X_{n_0-1}^*))=\rho_{2n_0}(\tilde X_{n_0-1}^*) . \nonumber
\end{eqnarray}
In general, for $i=1, \cdots, N-1$, we have
\begin{eqnarray}
\rho_{in_0}(X_{in_0}^*) &=& \rho_{in_0}(X_0^*) \nonumber \\
\rho_{in_0+1}(X_{in_0+1}^*) &=& \rho_{in_0}(\tilde X_1^*) \nonumber \\
&\cdots& \nonumber \\
\rho_{(i+1)n_0-1}(X_{(i+1)n_0-1}^*) &=& \rho_{in_0}(\tilde X_{n_0-1}^*) . \nonumber
\end{eqnarray}
Hence,
\begin{eqnarray}
X &=& X_0^*\cup \rho_1(X_1^*)\cup \cdots \cup \rho_{n_0-1}(X_{n_0-1}^*) \nonumber \\
&& \cup \rho_{n_0}(X_0^*)\cup \rho_{n_0}(\tilde X_1^*)\cup \cdots \cup \rho_{n_0}(\tilde X_{n_0-1}^*) \nonumber \\
&& \cdots \nonumber \\
&& \cup \rho_{(N-1)n_0}(X_0^*)\cup \rho_{(N-1)n_0}(\tilde X_1^*)\cup \cdots \cup \rho_{(N-1)n_0}(\tilde X_{n_0-1}^*) \nonumber
\end{eqnarray}
is obtained. Thus we have shown the following.
\newtheorem{pro}{Proposition}[section]
\begin{pro}
A characteristic matrix $X$ for a TB convolutional code $C$ generated by $G_N^{tb}$ is given by
\begin{eqnarray}
X &=& X_0^*\cup \rho_1(X_1^*)\cup \cdots \cup \rho_{n_0-1}(X_{n_0-1}^*) \nonumber \\
&& \cup \rho_{n_0}(X_0^*) \cup \rho_{n_0}(\tilde X_1^*)\cup \cdots \cup \rho_{n_0}(\tilde X_{n_0-1}^*) \nonumber \\
&& \cdots \nonumber \\
&& \cup \rho_{(N-1)n_0}(X_0^*) \cup \rho_{(N-1)n_0}(\tilde X_1^*)\cup \cdots \cup \rho_{(N-1)n_0}(\tilde X_{n_0-1}^*) .
\end{eqnarray}
\end{pro}
\newtheorem{cor}{Corollary}[section]
\begin{cor} If the relation
\begin{equation}
X_0^*\cup \rho_1(X_1^*)\cup \cdots \cup \rho_{n_0-1}(X_{n_0-1}^*) \subseteq X_0^* \cup \rho_{n_0}(X_0^*)
\end{equation}
holds, then a characteristic matrix $X$ is given by
\begin{equation}
X=X_0^* \cup \rho_{n_0}(X_0^*) \cup \rho_{2n_0}(X_0^*) \cup \cdots \cup \rho_{(N-1)n_0}(X_0^*) .
\end{equation}
\end{cor}
\par
\begin{IEEEproof}
From the assumption, we have
\begin{displaymath}
\rho_{n_0}(X_0^*)\cup \rho_{n_0}(\tilde X_1^*)\cup \cdots \cup \rho_{n_0}(\tilde X_{n_0-1}^*) \subseteq \rho_{n_0}(X_0^*) \cup \rho_{2n_0}(X_0^*) .
\end{displaymath}
Similarly, we have
\begin{eqnarray}
\rho_{2n_0}(X_0^*)\cup \rho_{2n_0}(\tilde X_1^*)\cup \cdots \cup \rho_{2n_0}(\tilde X_{n_0-1}^*) &\subseteq& \rho_{2n_0}(X_0^*) \cup \rho_{3n_0}(X_0^*) \nonumber \\
&\cdots& \nonumber \\
\rho_{(N-1)n_0}(X_0^*)\cup \rho_{(N-1)n_0}(\tilde X_1^*)\cup \cdots \cup \rho_{(N-1)n_0}(\tilde X_{n_0-1}^*) &\subseteq& \rho_{(N-1)n_0}(X_0^*) \cup \rho_{Nn_0}(X_0^*) \nonumber \\
&=& \rho_{(N-1)n_0}(X_0^*) \cup X_0^* . \nonumber
\end{eqnarray}
Then, from Proposition 3.1, it follows that
\begin{displaymath}
X=X_0^* \cup \rho_{n_0}(X_0^*) \cup \rho_{2n_0}(X_0^*) \cup \cdots \cup \rho_{(N-1)n_0}(X_0^*) .
\end{displaymath}
\end{IEEEproof}
\par
We remark that in many practical applications, a characteristic matrix for a TB convolutional code is obtained based on the above corollary.
\par
{\it Example 1:} Consider the TB convolutional code of section length $N=3$ defined by
\begin{eqnarray}
G(D) &=& (1+D, D, 1+D) \\
&=& (1, 0, 1)+(1, 1, 1)D \nonumber \\
&\stackrel{\triangle}{=}& G_0+G_1D .
\end{eqnarray}
The associated TBGM is given by
\begin{eqnarray}
G_3^{tb} &=& \left(
\begin{array}{ccc}
G_0 & G_1 & 0  \\
0 & G_0 & G_1  \\
G_1 & 0 & G_0
\end{array}
\right) \nonumber \\
&=& \left(
\begin{array}{cccccccccc}
1 & 0 & 1 & 1 & 1 & 1 & 0 & 0 & 0  \\
0 & 0 & 0 & 1 & 0 & 1 & 1 & 1 & 1 \\
1 & 1 & 1 & 0 & 0 & 0 & 1 & 0 & 1
\end{array}
\right) . \nonumber
\end{eqnarray}
In this case, we have
\begin{equation}
X_0^*=\left(
\begin{array}{ccccccccc}
\mbox{\boldmath $1$} & \mbox{\boldmath $0$} & \mbox{\boldmath $1$} & \mbox{\boldmath $1$} & \mbox{\boldmath $1$} & \mbox{\boldmath $1$} & 0 & 0 & 0 \\
0 & 0 & 0 & \mbox{\boldmath $1$} & \mbox{\boldmath $0$}& \mbox{\boldmath $1$} & \mbox{\boldmath $1$} & \mbox{\boldmath $1$} & \mbox{\boldmath $1$} \\
0 & \mbox{\boldmath $1$} & \mbox{\boldmath $0$} & \mbox{\boldmath $0$} & \mbox{\boldmath $1$} & \mbox{\boldmath $0$}& \mbox{\boldmath $0$} & \mbox{\boldmath $1$} & 0 \\
\end{array}
\right)
\begin{array}{c}
$(0, 5]$ \\
$(3, 8]$ \\
$(1, 7]$
\end{array}
\end{equation}
\begin{equation}
\tilde X_1^*=\left(
\begin{array}{ccccccccc}
\mbox{\boldmath $1$} & 0 & \mbox{\boldmath $1$} & \mbox{\boldmath $1$} & \mbox{\boldmath $1$} & \mbox{\boldmath $1$} & \mbox{\boldmath $0$} & \mbox{\boldmath $0$} & \mbox{\boldmath $0$} \\
0 & 0 & 0 & \mbox{\boldmath $1$} & \mbox{\boldmath $0$}& \mbox{\boldmath $1$} & \mbox{\boldmath $1$} & \mbox{\boldmath $1$} & \mbox{\boldmath $1$} \\
0 & \mbox{\boldmath $1$} & \mbox{\boldmath $0$} & \mbox{\boldmath $0$} & \mbox{\boldmath $1$} & \mbox{\boldmath $0$}& \mbox{\boldmath $0$} & \mbox{\boldmath $1$} & 0 \\
\end{array}
\right)
\begin{array}{l}
$(2, 0]$ \\
$(3, 8]$ \\
$(1, 7]$
\end{array}
\end{equation}
\begin{equation}
\tilde X_2^*=\left(
\begin{array}{ccccccccc}
\mbox{\boldmath $1$} & 0 & \mbox{\boldmath $1$} & \mbox{\boldmath $1$} & \mbox{\boldmath $1$} & \mbox{\boldmath $1$} & \mbox{\boldmath $0$} & \mbox{\boldmath $0$} & \mbox{\boldmath $0$} \\
0 & 0 & 0 & \mbox{\boldmath $1$} & \mbox{\boldmath $0$}& \mbox{\boldmath $1$} & \mbox{\boldmath $1$} & \mbox{\boldmath $1$} & \mbox{\boldmath $1$} \\
\mbox{\boldmath $0$} & \mbox{\boldmath $1$} & 0 & 0 & \mbox{\boldmath $1$} & \mbox{\boldmath $0$} & \mbox{\boldmath $0$} & \mbox{\boldmath $1$}& \mbox{\boldmath $0$} \\
\end{array}
\right)
\begin{array}{l}
$(2, 0]$ \\
$(3, 8]$ \\
$(4, 1]$ .
\end{array}
\end{equation}
By applying $\rho_3$ and $\rho_6$ to these matrices, a characteristic matrix $X$ is obtained as follows:
\begin{equation}
X=\left(
\begin{array}{ccccccccc}
\mbox{\boldmath $1$} & \mbox{\boldmath $0$} & \mbox{\boldmath $1$} & \mbox{\boldmath $1$} & \mbox{\boldmath $1$} & \mbox{\boldmath $1$} & 0 & 0 & 0 \\
0 & \mbox{\boldmath $1$} & \mbox{\boldmath $0$} & \mbox{\boldmath $0$} & \mbox{\boldmath $1$} & \mbox{\boldmath $0$}& \mbox{\boldmath $0$} & \mbox{\boldmath $1$} & 0 \\
\mbox{\boldmath $1$} & 0 & \mbox{\boldmath $1$} & \mbox{\boldmath $1$} & \mbox{\boldmath $1$} & \mbox{\boldmath $1$} & \mbox{\boldmath $0$} & \mbox{\boldmath $0$} & \mbox{\boldmath $0$} \\
0 & 0 & 0 & \mbox{\boldmath $1$} & \mbox{\boldmath $0$}& \mbox{\boldmath $1$} & \mbox{\boldmath $1$} & \mbox{\boldmath $1$} & \mbox{\boldmath $1$} \\
\mbox{\boldmath $0$} & \mbox{\boldmath $1$} & 0 & 0 & \mbox{\boldmath $1$} & \mbox{\boldmath $0$} & \mbox{\boldmath $0$} & \mbox{\boldmath $1$}& \mbox{\boldmath $0$} \\
\mbox{\boldmath $0$} & \mbox{\boldmath $0$} & \mbox{\boldmath $0$} & \mbox{\boldmath $1$} & 0 & \mbox{\boldmath $1$} & \mbox{\boldmath $1$} & \mbox{\boldmath $1$} & \mbox{\boldmath $1$} \\
\mbox{\boldmath $1$}& \mbox{\boldmath $1$} & \mbox{\boldmath $1$} & 0 & 0 & 0 & \mbox{\boldmath $1$} & \mbox{\boldmath $0$} & \mbox{\boldmath $1$} \\
\mbox{\boldmath $0$} & \mbox{\boldmath $1$} & \mbox{\boldmath $0$}& \mbox{\boldmath $0$} & \mbox{\boldmath $1$} & 0 & 0 & \mbox{\boldmath $1$} & \mbox{\boldmath $0$} \\
\mbox{\boldmath $1$} & \mbox{\boldmath $1$}& \mbox{\boldmath $1$} & \mbox{\boldmath $0$} & \mbox{\boldmath $0$} & \mbox{\boldmath $0$} & \mbox{\boldmath $1$} & 0 & \mbox{\boldmath $1$}
\end{array}
\right)
\begin{array}{l}
$(0, 5]$ \\
$(1, 7]$ \\
$(2, 0]$ \\
$(3, 8]$ \\
$(4, 1]$ \\
$(5, 3]$ \\
$(6, 2]$ \\
$(7, 4]$ \\
$(8, 6]$ .
\end{array}
\end{equation}
Note that the spans $(0, 5],~(1, 7],~(2, 0],~(3, 8],~(4, 1]$ are connected with $X_0^*\cup \rho_1(X_1^*)\cup \rho_2(X_2^*)$, whereas the spans $(0, 5],~(1, 7],~(3, 8],~(4, 1],~(6, 2]$ are connected with $X_0^* \cup \rho_3(X_0^*)$. Hence,
\begin{displaymath}
X_0^*\cup \rho_1(X_1^*)\cup \rho_2(X_2^*) \nsubseteq X_0^* \cup \rho_3(X_0^*) .
\end{displaymath}
We see that a characteristic matrix $X$ cannot be obtained simply by applying $\rho_3$ and $\rho_6$ to $X_0^*$.

\subsection{Structure of the Characteristic Span List}
Let $(X, T)$, where
\begin{displaymath}
X=\left(
\begin{array}{c}
x_1  \\
x_2  \\
\cdots \\
x_n
\end{array}
\right)
\end{displaymath}
\begin{displaymath}
T=\left(
\begin{array}{c}
(a_1, b_1]  \\
(a_2, b_2]  \\
\cdots \\
(a_n, b_n]
\end{array}
\right) ,
\end{displaymath}
be a characteristic matrix for $C$ with span list $T$, then $(\sigma_1(X), \sigma_1(T))$ is a characteristic matrix for $\sigma_1(C)$ with span list $\sigma_1(T)$~\cite[Remark III.9 (b)]{glu 111}, where
\begin{equation}
\sigma_1(X)\stackrel{\triangle}{=}\left(
\begin{array}{c}
\sigma_1(x_1)  \\
\sigma_1(x_2)  \\
\cdots \\
\sigma_1(x_n)
\end{array}
\right)
\end{equation}
\begin{equation}
\sigma_1(T)\stackrel{\triangle}{=}\left(
\begin{array}{c}
(a_1-1, b_1-1]  \\
(a_2-1, b_2-1]  \\
\cdots \\
(a_n-1, b_n-1]
\end{array}
\right) .
\end{equation}
Using repeatedly this relation, we see that $(\sigma_j(X), \sigma_j(T))$ is a characteristic matrix for $\sigma_j(C)$ with span list $\sigma_j(T)$. Consider a TB convolutional code $C$ generated by $G_N^{tb}$ and set $j$ to $n_0$. Since $\sigma_{n_0}(G_N^{tb})$ is equivalent to $G_N^{tb}$, $\sigma_{n_0}(C)=C$ holds. Thus we have the following.
\newtheorem{lem}{Lemma}[section]
\begin{lem}
Let $C$ be a TB convolutional code generated by $G_N^{tb}$. If $(X, T)$ is a characteristic matrix for $C$ with span list $T$, then $(\sigma_{n_0}(X), \sigma_{n_0}(T))$ is also a characteristic matrix for $C$ with span list $\sigma_{n_0}(T)$. Let
\begin{equation}
T=\{(a_l, b_l]:~l=1, 2, \cdots, n\} .
\end{equation}
Then $\sigma_{n_0}(T)$ is given by
\begin{equation}
\sigma_{n_0}(T)=\{(a_l-n_0, b_l-n_0]:~l=1, 2, \cdots, n\} .
\end{equation}
Since the characteristic span list is uniquely determined, $T$ and $\sigma_{n_0}(T)$ coincide up to ordering.
\end{lem}
\begin{pro}
The characteristic span list $T$ of a TB convolutional code $C$ generated by $G_N^{tb}$ consists of the set of basic spans
\begin{equation}
T_0\stackrel{\triangle}{=}\left\{
\begin{array}{l}
(0, b_0] \\
(1, b_1] \\
\cdots \\
(n_0-1, b_{n_0-1}]
\end{array} \right.
\end{equation}
and $\rho_{in_0}(T_0)~(i=1, 2, \cdots, N-1)$.
\end{pro}
\begin{IEEEproof}
Suppose that the spans in $T$ are sorted such that
\begin{displaymath}
T=\{(l, b_l]:~l=0, 1, \cdots, n-1\} .
\end{displaymath}
Then we have
\begin{displaymath}
\sigma_{n_0}(T)=\{(l-n_0, b_l-n_0]:~l=0, 1, \cdots, n-1\} .
\end{displaymath}
Here take notice of the following set of spans in $T$:
\begin{displaymath}
\left\{
\begin{array}{l}
(n_0, b_{n_0}] \\
(n_0+1, b_{n_0+1}] \\
\cdots \\
(2n_0-1, b_{2n_0-1}] .
\end{array} \right.
\end{displaymath}
In $\sigma_{n_0}(T)$, it is transformed to
\begin{displaymath}
\left\{
\begin{array}{l}
(0, b_{n_0}-n_0] \\
(1, b_{n_0+1}-n_0] \\
\cdots \\
(n_0-1, b_{2n_0-1}-n_0] .
\end{array} \right.
\end{displaymath}
Since $T$ and $\sigma_{n_0}(T)$ coincide up to ordering,
\begin{displaymath}
\left\{
\begin{array}{l}
(0, b_{n_0}-n_0]=(0, b_0] \\
(1, b_{n_0+1}-n_0]=(1, b_1] \\
\cdots \\
(n_0-1, b_{2n_0-1}-n_0]=(n_0-1, b_{n_0-1}]
\end{array} \right.
\end{displaymath}
holds. Hence, we have
\begin{displaymath}
\left\{
\begin{array}{l}
b_{n_0}=b_0+n_0 \\
b_{n_0+1}=b_1+n_0 \\
\cdots \\
b_{2n_0-1}=b_{n_0-1}+n_0 .
\end{array} \right.
\end{displaymath}
Similarly, the set of spans
\begin{displaymath}
\left\{
\begin{array}{l}
(2n_0, b_{2n_0}] \\
(2n_0+1, b_{2n_0+1}] \\
\cdots \\
(3n_0-1, b_{3n_0-1}]
\end{array} \right.
\end{displaymath}
is transformed to
\begin{displaymath}
\left\{
\begin{array}{l}
(n_0, b_{2n_0}-n_0] \\
(n_0+1, b_{2n_0+1}-n_0] \\
\cdots \\
(2n_0-1, b_{3n_0-1}-n_0] .
\end{array} \right.
\end{displaymath}
Then for the same reason, 
\begin{displaymath}
\left\{
\begin{array}{l}
(n_0, b_{2n_0}-n_0]=(n_0, b_{n_0}] \\
(n_0+1, b_{2n_0+1}-n_0]=(n_0+1, b_{n_0+1}] \\
\cdots \\
(2n_0-1, b_{3n_0-1}-n_0]=(2n_0-1, b_{2n_0-1}]
\end{array} \right.
\end{displaymath}
holds. Hence, we have
\begin{displaymath}
\left\{
\begin{array}{l}
b_{2n_0}=b_{n_0}+n_0=b_0+2n_0 \\
b_{2n_0+1}=b_{n_0+1}+n_0=b_1+2n_0 \\
\cdots \\
b_{3n_0-1}=b_{2n_0-1}+n_0=b_{n_0-1}+2n_0 .
\end{array} \right.
\end{displaymath}
Continuing the same argument, we have
\begin{displaymath}
\left\{
\begin{array}{l}
b_{in_0}=b_0+in_0 \\
b_{in_0+1}=b_1+in_0 \\
\cdots \\
b_{(i+1)n_0-1}=b_{n_0-1}+in_0
\end{array} \right.
\end{displaymath}
for $i=1, \cdots, N-1$.
\end{IEEEproof}
\par
{\it Example 2:} Consider the TB convolutional code of section length $N=3$ defined by the rate $R=2/3$ encoder
\begin{eqnarray}
G(D) &=& \left(
\begin{array}{ccc}
1+D & D & 1+D \\
D & 1 & 1
\end{array}
\right) \\
&=& \left(
\begin{array}{ccc}
1 & 0 & 1 \\
0 & 1 & 1
\end{array}
\right)+\left(
\begin{array}{ccc}
1 & 1 & 1 \\
1 & 0 & 0
\end{array}
\right)D \nonumber \\
&\stackrel{\triangle}{=}& G_0+G_1D .
\end{eqnarray}
Using the associated TBGM, i.e., 
\begin{displaymath}
G_3^{tb}=\left(
\begin{array}{ccc}
G_0 & G_1 & 0  \\
0 & G_0 & G_1  \\
G_1 & 0 & G_0
\end{array}
\right) ,
\end{displaymath}
a charactreristic matrix $X$ is computed as follows:
\begin{equation}
X=\left(
\begin{array}{ccccccccc}
\mbox{\boldmath $1$} & \mbox{\boldmath $0$} & \mbox{\boldmath $0$} & \mbox{\boldmath $1$}& \mbox{\boldmath $1$} & 0 & 0 & 0 & 0 \\
0 & \mbox{\boldmath $1$} & \mbox{\boldmath $1$} & \mbox{\boldmath $1$} & 0 & 0 & 0 & 0 & 0 \\
0 & 0 & \mbox{\boldmath $1$} & \mbox{\boldmath $0$} & \mbox{\boldmath $0$} & \mbox{\boldmath $1$}& 0 & 0 & 0 \\
0 & 0 & 0 & \mbox{\boldmath $1$} & \mbox{\boldmath $0$} & \mbox{\boldmath $0$} & \mbox{\boldmath $1$}& \mbox{\boldmath $1$} & 0 \\
0 & 0 & 0 & 0 & \mbox{\boldmath $1$} & \mbox{\boldmath $1$} & \mbox{\boldmath $1$} & 0 & 0 \\
0 & 0 & 0 & 0 & 0 &\mbox{\boldmath $1$} & \mbox{\boldmath $0$} & \mbox{\boldmath $0$} & \mbox{\boldmath $1$} \\
\mbox{\boldmath $1$}& \mbox{\boldmath $1$} & 0 & 0 & 0 & 0 & \mbox{\boldmath $1$} & \mbox{\boldmath $0$} & \mbox{\boldmath $0$} \\
\mbox{\boldmath $1$} & 0 & 0 & 0 & 0 & 0 & 0 & \mbox{\boldmath $1$} & \mbox{\boldmath $1$} \\
\mbox{\boldmath $0$} & \mbox{\boldmath $0$}& \mbox{\boldmath $1$} & 0 & 0 & 0 & 0 & 0 & \mbox{\boldmath $1$}
\end{array}
\right)
\begin{array}{l}
$(0, 4]$ \\
$(1, 3]$ \\
$(2, 5]$ \\
$(3, 7]$ \\
$(4, 6]$ \\
$(5, 8]$ \\
$(6, 1]$ \\
$(7, 0]$ \\
$(8, 2]$ .
\end{array}
\end{equation}
We see that the characteristic span list $T$ consists of the set of basic spans
\begin{equation}
T_0=\left\{
\begin{array}{l}
(0, 4] \\
(1, 3] \\
(2, 5]
\end{array} \right.
\end{equation}
and its right cyclic shifts by $3$ and $6$ positions.

\subsection{Counting Characteristic Matrices}
Recall the definition of a characteristic matrix $X$ for a given code $C$, i.e.,
\begin{displaymath}
X\stackrel{\triangle}{=}X_0^* \cup \rho_1(X_1^*)\cup \cdots \cup \rho_{n-1}(X_{n-1}^*) ,
\end{displaymath}
where $X_j^*$ is a basis in minimal-span form for the code $C_j=\sigma_j(C)$. Note that $X_j^*$ is not necessarily unique. Hence, $X$ is not uniquely determined~\cite{glu 111}. With respect to this subject, Weaver~\cite{wea 12} discussed the relationship between the characteristic span list of $C$ and the number of characteristic matrices for $C$.
\par
Let $T=\{(a_l, b_l]:~l=1, 2, \cdots, n\}$ be the characteristic span list of $C$. Define the set $\Theta_l$ as follows~\cite{wea 12}:
\begin{equation}
\Theta_l\stackrel{\triangle}{=}\{r:~(a_r, b_r]\subsetneq (a_l, b_l]\} .
\end{equation}
$\vert \Theta_l \vert$ represents the number of spans (in $T$) included in a specified span $(a_l, b_l]$. Weaver~\cite{wea 12} proved the following.
\begin{lem}[Weaver~\cite{wea 12}]
Let $(a_l, b_l]$ be a characteristic span of $C$. Then there exist $2^{\vert \Theta_l \vert}$ characteristic generators for $C$ having this span.
\end{lem}
\par
This fact is derived from the next observation:
\par
Let $(a_r, b_r]\subsetneq (a_l, b_l]$ and consider two characteristic generators $x_r$ and $x_l$ with spans $(a_r, b_r]$ and $(a_l, b_l]$, respectively. Then $x_l+x_r$ is also a characteristic generator with span $(a_l, b_l]$.
\par
Consider a TB convolutional code $C$ generated by $G_N^{tb}$. We have already shown that the characteristic span list $T$ of $C$ consists of the set of basic spans
\begin{displaymath}
T_0=\left\{
\begin{array}{l}
(0, b_0] \\
(1, b_1] \\
\cdots \\
(n_0-1, b_{n_0-1}]
\end{array} \right.
\end{displaymath}
and $\rho_{in_0}(T_0)~(i=1, 2, \cdots, N-1)$. Hence, it suffices to consider the spans in $T_0$ for the purpose of counting the number of characteristic matrices. Define $\Theta_i~(1 \leq i \leq n_0)$ as follows:
\begin{equation}
\left\{
\begin{array}{l}
\Theta_1\stackrel{\triangle}{=}\{r:~(a_r, b_r]\subsetneq (0, b_0]\} \\
\Theta_2\stackrel{\triangle}{=}\{r:~(a_r, b_r]\subsetneq (1, b_1]\} \\
\cdots \\
\Theta_{n_0}\stackrel{\triangle}{=}\{r:~(a_r, b_r]\subsetneq (n_0-1, b_{n_0-1}]\} .
\end{array} \right.
\end{equation}
Also, let
\begin{equation}
\theta_i\stackrel{\triangle}{=}\vert \Theta_i \vert ~(i=1, 2, \cdots, n_0)
\end{equation}
\begin{equation}
\theta\stackrel{\triangle}{=}\theta_1 + \theta_2 + \cdots + \theta_{n_0} .
\end{equation}
Then we have the following:
\begin{itemize}
\item There exist $2^{\theta_1}$ characteristic generators having span $(0, b_0]$.
\item There exist $2^{\theta_2}$ characteristic generators having span $(1, b_1]$.
\item[] $\cdots$
\item There exist $2^{\theta_{n_0}}$ characteristic generators having span $(n_0-1, b_{n_0-1}]$.
\end{itemize}
As a result, the degree of freedom related to the spans in $T_0$ is given by
\begin{eqnarray}
\lefteqn{2^{\theta_1} \times 2^{\theta_2} \times \cdots \times 2^{\theta_{n_0}}} \nonumber \\
&& =2^{\theta_1+\theta_2+\cdots +\theta_{n_0}} \nonumber \\
&& =2^{\theta} .
\end{eqnarray}
Since this degree of freedom is common to other $(N-1)$ blocks of spans in $T$, the overall degree of freedom related to $T$ becomes
\begin{equation}
(2^{\theta})^N=2^{\theta N} .
\end{equation}
Thus we have shown the following.
\begin{pro}
Let $C$ be a TB convolutional code generated by $G_N^{tb}$. Let $\theta_i~(1 \leq i \leq n_0)$ and $\theta$ be as above. Then there exist $2^{\theta N}$ characteristic matrices for $C$.
\end{pro}
\par
{\it Example 2 (Continued):} Take notice of the first three rows of the characteristic matrix $X$. We have
\begin{displaymath}
\left\{
\begin{array}{l}
\Theta_1=\{2\} \\
\Theta_2=\phi \\
\Theta_3=\phi .
\end{array} \right.
\end{displaymath}
Hence, $\theta=1+0+0=1$ and there exist $2^{1 \times 3}=8$ characteristic matrices.

\subsection{Span Lengths of Characteristic Generators}
Let $[x]=(a, b]$ be a span of a codeword $x$. Then the span length of $x$ is defined by $\vert [a, b] \vert$, i.e., the number of elements in the closed interval $[a, b]$. When a span alone is referred to without specifying the accompanied codeword, we use the term {\it the span length of a span $(a, b]$}. Let $T$ be the characteristic span list of a TB convolutional code $C$ generated by $G_N^{tb}$. Suppose that the spans in $T$ are sorted such that
\begin{displaymath}
T=\{(l, b_l]:~l=0, 1, \cdots, n-1\} .
\end{displaymath}
Then by~\cite[Theorem 5.10]{koe 03},
\begin{equation}
\vert (0, b_0] \vert +\vert (1, b_1] \vert+ \cdots+ \vert (n-1, b_{n-1}]\vert=n(n-k)
\end{equation}
holds. Due to the structure of $T$ (see Proposition 3.2), the left-hand side of the above equality becomes
\begin{eqnarray}
\lefteqn{N(\vert (0, b_0] \vert +\vert (1, b_1] \vert+ \cdots+ \vert (n_0-1, b_{n_0-1}]\vert)} \nonumber \\
&& =N((\ell_1-1)+(\ell_2-1)+\cdots +(\ell_{n_0}-1)) \nonumber \\
&& =N((\ell_1+\ell_2+\cdots +\ell_{n_0})-n_0) \nonumber \\
&& =N(\ell-n_0) , \nonumber
\end{eqnarray}
where
\begin{equation}
\left\{
\begin{array}{l}
\ell_1\stackrel{\triangle}{=}\vert [0, b_0]\vert \\
\ell_2\stackrel{\triangle}{=}\vert [1, b_1]\vert  \\
\cdots \\
\ell_{n_0}\stackrel{\triangle}{=}\vert [n_0-1, b_{n_0-1}]\vert
\end{array} \right.
\end{equation}
\begin{equation}
\ell \stackrel{\triangle}{=}\ell_1+\ell_2+\cdots +\ell_{n_0} .
\end{equation}
In the derivation, we also used the relation
\begin{displaymath}
\ell_i=\vert[i-1, b_{i-1}]\vert=\vert(i-1, b_{i-1}]\vert+1~(i=1, 2, \cdots, n_0) .
\end{displaymath}
Replacing $n$ and $k$ by $n_0N$ and $k_0N$, respectively, the above equality reduces to
\begin{eqnarray}
N(\ell-n_0) &=& n_0N(n_0N-k_0N) \nonumber \\
\ell-n_0 &=& n_0((n_0-k_0)N) \nonumber \\
\ell &=& n_0((n_0-k_0)N+1) . \nonumber
\end{eqnarray}
Thus we have shown the following.
\begin{pro}
Let $T$ be the characteristic span list of a TB convolutional code $C$ generated by $G_N^{tb}$. Denote by $T_0$ the set of basic spans in $T$. Then the sum $\ell$ of span lengths of spans in $T_0$ is given by
\begin{equation}
\ell=n_0((n_0-k_0)N+1) .
\end{equation}
\end{pro}
\par
{\it Example 2 (Continued):} We have
\begin{displaymath}
\left\{
\begin{array}{l}
\ell_1=\vert [0, 4]\vert=5 \\
\ell_2=\vert [1, 3]\vert=3 \\
\ell_3=\vert [2, 5]\vert=4
\end{array} \right.
\end{displaymath}
\begin{displaymath}
\ell=\ell_1+\ell_2+\ell_3=5+3+4=12 .
\end{displaymath}
Also, we have
\begin{displaymath}
n_0((n_0-k_0)N+1)=3((3-2) \times 3+1)=12 .
\end{displaymath}


\section{Transformations of $G(D)$ and the Corresponding TBGM's}
In this section, we discuss the relationship between transformations of a generator matrix $G(D)$ and the corresponding TBGM's ($G_N^{tb}\mbox{'}s$). We consider the following transformations of $G(D)$:
\begin{itemize}
\item[a)] Dividing the $j$th column by $D^p$.
\item[b)] Multiplying the $j$th column by $D^q$.
\item[c)] Adding the $i$th row multiplied by $D^q$ to the $j$th row.
\item[d)] Implicit transformations.
\end{itemize}
\par
In the next section, we will see that these transformations play an essential role in trellis reduction for TB convolutional codes.

\subsection{Dividing a Column of $G(D)$ by $D^p$}
Suppose that the $j$th column of $G(D)$ has a monomial factor $D^p~(1 \leq p \leq L)$. We can assume without loss of generality that $j=1$ and $p=1$. Hence, $G(D)$ has the form
\begin{equation}
G(D)=\left(
\begin{array}{cccc}
g_{1, 1}'(D)D & g_{1, 2}(D) & \ldots & g_{1, n_0}(D)  \\
g_{2, 1}'(D)D & g_{2, 2}(D) & \ldots & g_{2, n_0}(D)  \\
\ldots & \ldots & \ldots & \ldots \\
g_{k_0, 1}'(D)D & g_{k_0, 2}(D) & \ldots & g_{k_0, n_0}(D)
\end{array}
\right) .
\end{equation}
Let
\begin{eqnarray}
G(D) &=& \left(
\begin{array}{cccc}
g_{1, 1}^{(0)} & g_{1, 2}^{(0)} & \ldots & g_{1, n_0}^{(0)}  \\
g_{2, 1}^{(0)} & g_{2, 2}^{(0)} & \ldots & g_{2, n_0}^{(0)}  \\
\ldots & \ldots & \ldots & \ldots \\
g_{k_0, 1}^{(0)} & g_{k_0, 2}^{(0)} & \ldots & g_{k_0, n_0}^{(0)}
\end{array}
\right)+\left(
\begin{array}{cccc}
g_{1, 1}^{(1)} & g_{1, 2}^{(1)} & \ldots & g_{1, n_0}^{(1)}  \\
g_{2, 1}^{(1)} & g_{2, 2}^{(1)} & \ldots & g_{2, n_0}^{(1)}  \\
\ldots & \ldots & \ldots & \ldots \\
g_{k_0, 1}^{(1)} & g_{k_0, 2}^{(1)} & \ldots & g_{k_0, n_0}^{(1)}
\end{array}
\right)D \nonumber \\
&& +\left(
\begin{array}{cccc}
g_{1, 1}^{(2)} & g_{1, 2}^{(2)} & \ldots & g_{1, n_0}^{(2)}  \\
g_{2, 1}^{(2)} & g_{2, 2}^{(2)} & \ldots & g_{2, n_0}^{(2)}  \\
\ldots & \ldots & \ldots & \ldots \\
g_{k_0, 1}^{(2)} & g_{k_0, 2}^{(2)} & \ldots & g_{k_0, n_0}^{(2)}
\end{array}
\right)D^2+\cdots \nonumber \\
&& +\left(
\begin{array}{cccc}
g_{1, 1}^{(L)} & g_{1, 2}^{(L)} & \ldots & g_{1, n_0}^{(L)}  \\
g_{2, 1}^{(L)} & g_{2, 2}^{(L)} & \ldots & g_{2, n_0}^{(L)}  \\
\ldots & \ldots & \ldots & \ldots \\
g_{k_0, 1}^{(L)} & g_{k_0, 2}^{(L)} & \ldots & g_{k_0, n_0}^{(L)}
\end{array}
\right)D^L \nonumber \\
&\stackrel{\triangle}{=}& G_0+G_1D+\cdots +G_{L-1}D^{L-1}+G_LD^L
\end{eqnarray}
be the polynomial expansion of $G(D)$. Comparing the $(i, 1)~(1 \leq i \leq k_0)$ entries, we have
\begin{displaymath}
\left\{
\begin{array}{l}
g_{1, 1}'(D)D=g_{1, 1}^{(0)}+g_{1, 1}^{(1)}D+g_{1, 1}^{(2)}D^2+\cdots +g_{1, 1}^{(L)}D^L \\
g_{2, 1}'(D)D=g_{2, 1}^{(0)}+g_{2, 1}^{(1)}D+g_{2, 1}^{(2)}D^2+\cdots +g_{2, 1}^{(L)}D^L \\
\cdots \\
g_{k_0, 1}'(D)D=g_{k_0, 1}^{(0)}+g_{k_0, 1}^{(1)}D+g_{k_0, 1}^{(2)}D^2+\cdots +g_{k_0, 1}^{(L)}D^L .
\end{array} \right.
\end{displaymath}
By these equations, we have
\begin{equation}
\left\{
\begin{array}{l}
g_{1, 1}^{(0)}=0 \\
g_{2, 1}^{(0)}=0 \\
\cdots \\
g_{k_0, 1}^{(0)}=0
\end{array} \right.
\end{equation}
\begin{equation}
\left\{
\begin{array}{l}
g_{1, 1}'(D)=g_{1, 1}^{(1)}+g_{1, 1}^{(2)}D+\cdots +g_{1, 1}^{(L)}D^{L-1}+0 \cdot D^L \\
g_{2, 1}'(D)=g_{2, 1}^{(1)}+g_{2, 1}^{(2)}D+\cdots +g_{2, 1}^{(L)}D^{L-1}+0 \cdot D^L \\
\cdots \\
g_{k_0, 1}'(D)=g_{k_0, 1}^{(1)}+g_{k_0, 1}^{(2)}D+\cdots +g_{k_0, 1}^{(L)}D^{L-1}+0 \cdot D^L .
\end{array} \right.
\end{equation}
Dividing the first column of $G(D)$ by $D$, let the resulting matrix be $G'(D)$. Then $G'(D)$ has the polynomial expansion:
\begin{eqnarray}
G'(D) &=& \left(
\begin{array}{cccc}
g_{1, 1}'(D) & g_{1, 2}(D) & \ldots & g_{1, n_0}(D)  \\
g_{2, 1}'(D) & g_{2, 2}(D) & \ldots & g_{2, n_0}(D)  \\
\ldots & \ldots & \ldots & \ldots \\
g_{k_0, 1}'(D) & g_{k_0, 2}(D) & \ldots & g_{k_0, n_0}(D)
\end{array}
\right) \\
&=& \left(
\begin{array}{cccc}
g_{1, 1}^{(1)} & g_{1, 2}^{(0)} & \ldots & g_{1, n_0}^{(0)}  \\
g_{2, 1}^{(1)} & g_{2, 2}^{(0)} & \ldots & g_{2, n_0}^{(0)}  \\
\ldots & \ldots & \ldots & \ldots \\
g_{k_0, 1}^{(1)} & g_{k_0, 2}^{(0)} & \ldots & g_{k_0, n_0}^{(0)}
\end{array}
\right)+\left(
\begin{array}{cccc}
g_{1, 1}^{(2)} & g_{1, 2}^{(1)} & \ldots & g_{1, n_0}^{(1)}  \\
g_{2, 1}^{(2)} & g_{2, 2}^{(1)} & \ldots & g_{2, n_0}^{(1)}  \\
\ldots & \ldots & \ldots & \ldots \\
g_{k_0, 1}^{(2)} & g_{k_0, 2}^{(1)} & \ldots & g_{k_0, n_0}^{(1)}
\end{array}
\right)D \nonumber \\
&& +\left(
\begin{array}{cccc}
g_{1, 1}^{(3)} & g_{1, 2}^{(2)} & \ldots & g_{1, n_0}^{(2)}  \\
g_{2, 1}^{(3)} & g_{2, 2}^{(2)} & \ldots & g_{2, n_0}^{(2)}  \\
\ldots & \ldots & \ldots & \ldots \\
g_{k_0, 1}^{(3)} & g_{k_0, 2}^{(2)} & \ldots & g_{k_0, n_0}^{(2)}
\end{array}
\right)D^2+\cdots \nonumber \\
&& +\left(
\begin{array}{cccc}
g_{1, 1}^{(L)} & g_{1, 2}^{(L-1)} & \ldots & g_{1, n_0}^{(L-1)}  \\
g_{2, 1}^{(L)} & g_{2, 2}^{(L-1)} & \ldots & g_{2, n_0}^{(L-1)}  \\
\ldots & \ldots & \ldots & \ldots \\
g_{k_0, 1}^{(L)} & g_{k_0, 2}^{(L-1)} & \ldots & g_{k_0, n_0}^{(L-1)}
\end{array}
\right)D^{L-1}+\left(
\begin{array}{cccc}
0 & g_{1, 2}^{(L)} & \ldots & g_{1, n_0}^{(L)}  \\
0 & g_{2, 2}^{(L)} & \ldots & g_{2, n_0}^{(L)}  \\
\ldots & \ldots & \ldots & \ldots \\
0 & g_{k_0, 2}^{(L)} & \ldots & g_{k_0, n_0}^{(L)}
\end{array}
\right)D^L \nonumber \\
&\stackrel{\triangle}{=}& G_0'+G_1'D+\cdots +G_{L-1}'D^{L-1}+G_L'D^L .
\end{eqnarray}
\par
Consider the TBGM associated with $G'(D)$, denoted by $G_N'^{tb}$, where
\begin{equation}
\arraycolsep=1pt
G_N'^{tb}=\left(
\arraycolsep=1pt
\begin{array}{cccccccc}
G_0' & G_1' & \scriptstyle{\ldots} & G_{L-1}' & G_L' &  &  &  \\
 & G_0' & \scriptstyle{\ldots} & \scriptstyle{\ldots} & G_{L-1}' & G_L' &  &  \\
 &  & \scriptstyle{\ldots} & \scriptstyle{\ldots} & \scriptstyle{\ldots} & \scriptstyle{\ldots} & \scriptstyle{\ldots} &  \\
 &  &  & G_0' & G_1' & \scriptstyle{\ldots} & \scriptstyle{\ldots} & G_L' \\
G_L' &  &  &  & G_0' & G_1' & \scriptstyle{\ldots} & G_{L-1}' \\
G_{L-1}' & G_L' &  &  &  & G_0' & \scriptstyle{\ldots} & \scriptstyle{\ldots} \\
\scriptstyle{\ldots} & \scriptstyle{\ldots} & \scriptstyle{\ldots} &  &  &  & \scriptstyle{\ldots} & G_1' \\
G_1' & G_2' & \scriptstyle{\ldots} & G_L' &  &  &  & G_0'
\end{array}
\right) .
\end{equation}
Note that both $G_N^{tb}$ and $G_N'^{tb}$ can be regarded as matrices having $N$ blocks of $n_0$ columns. Then in view of the entries of $G_i'~(0 \leq i \leq L)$ and the relation
\begin{displaymath}
\left\{
\begin{array}{l}
g_{1, 1}^{(0)}=0 \\
g_{2, 1}^{(0)}=0 \\
\cdots \\
g_{k_0, 1}^{(0)}=0 ,
\end{array} \right.
\end{displaymath}
$G_N'^{tb}$ is obtained from $G_N^{tb}$ by cyclically shifting the first column of each block to the left by $n_0$ positions. Thus we have the following.
\begin{pro}
Regard $G_N^{tb}$ as a matrix having $N$ blocks of $n_0$ columns. Suppose that the $j$th column of $G(D)$ has a monomial factor $D^p~(1 \leq p \leq L)$. Then dividing the $j$th column of $G(D)$ by $D^p$ is equivalent to cyclically shifting the $j$th column of each block of $G_N^{tb}$ to the left by $pn_0$ positions.
\end{pro}
\par
Let $C$ be a TB convolutional code of section length $N$ defined by $G(D)$. Note that each codeword in $C$ consists of $N$ blocks of $n_0$ components. Here let us cyclically shift the $j$th component of each block to the left by $pn_0$ positions. Denote by $C'$ the set of resulting (modified) codewords. We have already shown that $G_N'^{tb}$ is obtained from $G_N^{tb}$ by cyclically shifting the $j$th column of each block to the left by $pn_0$ positions. Hence, $C'$ is generated by $G_N'^{tb}$. In words, $C'$ is represented as a TB convolutional code defined by $G'(D)$.

\subsection{Multiplying a Column of $G(D)$ by $D^q$}
Consider multiplication of the $j$th column of $G(D)$ by $D^q$, where $q+L+1 \leq N$. In the following, we assume without loss of generality that $j=1$C$q=1~(L+2 \leq N)$. Hence, the resulting matrix $G'(D)$ has the form
\begin{eqnarray}
G'(D) &=& \left(
\begin{array}{cccc}
g_{1, 1}(D)D & g_{1, 2}(D) & \ldots & g_{1, n_0}(D)  \\
g_{2, 1}(D)D & g_{2, 2}(D) & \ldots & g_{2, n_0}(D)  \\
\ldots & \ldots & \ldots & \ldots \\
g_{k_0, 1}(D)D & g_{k_0, 2}(D) & \ldots & g_{k_0, n_0}(D)
\end{array}
\right) \\
&\stackrel{\triangle}{=}& \left(
\begin{array}{cccc}
g_{1, 1}'(D) & g_{1, 2}(D) & \ldots & g_{1, n_0}(D)  \\
g_{2, 1}'(D) & g_{2, 2}(D) & \ldots & g_{2, n_0}(D)  \\
\ldots & \ldots & \ldots & \ldots \\
g_{k_0, 1}'(D) & g_{k_0, 2}(D) & \ldots & g_{k_0, n_0}(D)
\end{array}
\right) .
\end{eqnarray}
Then we have
\begin{eqnarray}
g_{1, 1}'(D) &=& g_{1, 1}(D)D \nonumber \\
&=& (g_{1, 1}^{(0)}+g_{1, 1}^{(1)}D+\cdots +g_{1, 1}^{(L)}D^L)D \nonumber \\
&=& g_{1, 1}^{(0)}D+g_{1, 1}^{(1)}D^2+\cdots +g_{1, 1}^{(L-1)}D^L+g_{1, 1}^{(L)}D^{L+1} \\
&\cdots& \nonumber \\
g_{k_0, 1}'(D) &=& g_{k_0, 1}(D)D \nonumber \\
&=& (g_{k_0, 1}^{(0)}+g_{k_0, 1}^{(1)}D+\cdots +g_{k_0, 1}^{(L)}D^L)D \nonumber \\
&=& g_{k_0, 1}^{(0)}D+g_{k_0, 1}^{(1)}D^2+\cdots +g_{k_0, 1}^{(L-1)}D^L+g_{k_0, 1}^{(L)}D^{L+1} .
\end{eqnarray}
Accordingly, the polynomial expansion of $G'(D)$ becomes
\begin{eqnarray}
G'(D) &=& \left(
\begin{array}{cccc}
0 & g_{1, 2}^{(0)} & \ldots & g_{1, n_0}^{(0)}  \\
0 & g_{2, 2}^{(0)} & \ldots & g_{2, n_0}^{(0)}  \\
\ldots & \ldots & \ldots & \ldots \\
0 & g_{k_0, 2}^{(0)} & \ldots & g_{k_0, n_0}^{(0)}
\end{array}
\right)+\left(
\begin{array}{cccc}
g_{1, 1}^{(0)} & g_{1, 2}^{(1)} & \ldots & g_{1, n_0}^{(1)}  \\
g_{2, 1}^{(0)} & g_{2, 2}^{(1)} & \ldots & g_{2, n_0}^{(1)}  \\
\ldots & \ldots & \ldots & \ldots \\
g_{k_0, 1}^{(0)} & g_{k_0, 2}^{(1)} & \ldots & g_{k_0, n_0}^{(1)}
\end{array}
\right)D \nonumber \\
&& +\left(
\begin{array}{cccc}
g_{1, 1}^{(1)} & g_{1, 2}^{(2)} & \ldots & g_{1, n_0}^{(2)}  \\
g_{2, 1}^{(1)} & g_{2, 2}^{(2)} & \ldots & g_{2, n_0}^{(2)}  \\
\ldots & \ldots & \ldots & \ldots \\
g_{k_0, 1}^{(1)} & g_{k_0, 2}^{(2)} & \ldots & g_{k_0, n_0}^{(2)}
\end{array}
\right)D^2+\cdots \nonumber \\
&& +\left(
\begin{array}{cccc}
g_{1, 1}^{(L-1)} & g_{1, 2}^{(L)} & \ldots & g_{1, n_0}^{(L)}  \\
g_{2, 1}^{(L-1)} & g_{2, 2}^{(L)} & \ldots & g_{2, n_0}^{(L)}  \\
\ldots & \ldots & \ldots & \ldots \\
g_{k_0, 1}^{(L-1)} & g_{k_0, 2}^{(L)} & \ldots & g_{k_0, n_0}^{(L)}
\end{array}
\right)D^L+\left(
\begin{array}{cccc}
g_{1, 1}^{(L)} & 0 & \ldots & 0  \\
g_{2, 1}^{(L)} & 0 & \ldots & 0  \\
\ldots & \ldots & \ldots & \ldots \\
g_{k_0, 1}^{(L)} & 0 & \ldots & 0
\end{array}
\right)D^{L+1} \nonumber \\
&\stackrel{\triangle}{=}& G_0'+G_1'D+\cdots +G_L'D^L+G_{L+1}'D^{L+1} .
\end{eqnarray}
\par
Consider the TBGM associated with $G'(D)$, denoted by $G_N'^{tb}$, where
\begin{equation}
\arraycolsep=1pt
G_N'^{tb}=\left(
\arraycolsep=1pt
\begin{array}{ccccccccc}
G_0' & G_1' & \scriptstyle{\ldots} & G_{L-1}' & G_L' & G_{L+1}' &  &  &  \\
 & G_0' & \scriptstyle{\ldots} & \scriptstyle{\ldots} & G_{L-1}' & G_L' & G_{L+1}' &  &  \\
 &  & \scriptstyle{\ldots} & \scriptstyle{\ldots} & \scriptstyle{\ldots} & \scriptstyle{\ldots} & \scriptstyle{\ldots} & \scriptstyle{\ldots} &  \\
 &  &  & G_0' & G_1' & \scriptstyle{\ldots} & \scriptstyle{\ldots} & G_L' & G_{L+1}' \\
G_{L+1}' &  &  &  & G_0' & G_1' & \scriptstyle{\ldots} & \scriptstyle{\ldots} & G_L' \\
G_L' & G_{L+1}' &  &  &  & G_0' & G_1' & \scriptstyle{\ldots} & G_{L-1}' \\
G_{L-1}' & G_L' & G_{L+1}' &  &  &  & G_0' & \scriptstyle{\ldots} & \scriptstyle{\ldots} \\
\scriptstyle{\ldots} & \scriptstyle{\ldots} & \scriptstyle{\ldots} & \scriptstyle{\ldots} &  &  &  & \scriptstyle{\ldots} & G_1' \\
G_1' & G_2' & \scriptstyle{\ldots} & G_L' & G_{L+1}' &  &  &  & G_0'
\end{array}
\right) . 
\end{equation}
Note that $G_N^{tb}$ and $G_N'^{tb}$ consist of $N$ blocks of $n_0$ columns as above. In view of the entries of $G_i'~(0 \leq i \leq L+1)$, we see that $G_N'^{tb}$ is obtained from $G_N^{tb}$ by cyclically shifting the first column of each block to the right by $n_0$ positions. Thus we have the following.
\begin{pro}
Regard $G_N^{tb}$ as a matrix having $N$ blocks of $n_0$ columns. Suppose that $q+L+1 \leq N$. Then multiplying the $j$th column of $G(D)$ by $D^q$ is equivalent to cyclically shifting the $j$th column of each block of $G_N^{tb}$ to the right by $qn_0$ positions.
\end{pro}
\par
{\it Remark:} In order for $G_N'^{tb}$ to be defined, the condition $q+L+1 \leq N$ is required.
\par
Let $C$ be a TB convolutional code of section length $N$ with generator matrix $G(D)$. Let $C'$ be as in the previous section. In this case, however, the $j$th component of each block is cyclically shifted to the right by $qn_0$ positions. We have shown that $G_N'^{tb}$ is obtained from $G_N^{tb}$ by cyclically shifting the $j$th column of each block to the right by $qn_0$ positions. Hence, $C'$ is generated by $G_N'^{tb}$. In words, $C'$ is represented as a TB convolutional code defined by $G'(D)$.

\subsection{$g_j(D) \leftarrow g_j(D)+D^qg_i(D)$}
Consider addition of the $i$th row $g_i(D)$ multiplied by $D^q$ to the $j$th row $g_j(D)$, denoted by $g_j(D) \leftarrow g_j(D)+D^qg_i(D)$, where $q+L+1 \leq N$. In the following, we assume without loss of generality that $i=1$ and $j=2$. Let the first row of $G(D)$ be
\begin{displaymath}
g_1(D)=(g_{1, 1}(D), g_{1, 2}(D), \cdots, g_{1, n_0}(D)) .
\end{displaymath}
Also, let
\begin{eqnarray}
g_1(D) &=& (g_{1, 1}^{(0)}, g_{1, 2}^{(0)}, \cdots, g_{1, n_0}^{(0)})+(g_{1, 1}^{(1)}, g_{1, 2}^{(1)}, \cdots, g_{1, n_0}^{(1)})D+ \cdots \nonumber \\
&& +(g_{1, 1}^{(L-1)}, g_{1, 2}^{(L-1)}, \cdots, g_{1, n_0}^{(L-1)})D^{L-1}+(g_{1, 1}^{(L)}, g_{1, 2}^{(L)}, \cdots, g_{1, n_0}^{(L)})D^L \nonumber \\
&\stackrel{\triangle}{=}& g_0+g_1D+ \cdots +g_{L-1}D^{L-1}+g_LD^L \nonumber
\end{eqnarray}
be the polynomial expansion, where the size of $g_i~(0 \leq i \leq L)$ is $1 \times n_0$. Then the polynomial expansion of $g_1(D)D^q$ becomes
\begin{displaymath}
g_1(D)D^q=g_0D^q+g_1D^{q+1}+ \cdots +g_{L-1}D^{L+q-1}+g_LD^{L+q} .
\end{displaymath}
Note that the first row of $G_N^{tb}$ is expressed as
\begin{displaymath}
\left(\underbrace{g_0, g_1, \cdots, g_{L-1}, g_L, 0, \cdots, 0}_{N} \right) .
\end{displaymath}
Hence, its right cyclic shift by $qn_0$ positions, i.e., 
\begin{displaymath}
\left(\underbrace{\overbrace{0, \cdots, 0}^{q}, \overbrace{g_0, g_1, \cdots, g_{L-1}, g_L}^{L+1}, 0, \cdots, 0}_{N} \right)
\end{displaymath}
coincides with the $(qk_0+1)$th row of $G_N^{tb}$. That is, $g_2(D) \leftarrow g_2(D)+D^qg_1(D)$ corresponds to addition of the $(qk_0+1)$th row to the second row within the matrix $G_N^{tb}$. Note that this is an elementary row operation. Thus we have the following.
\begin{pro}
Suppose that $q+L+1 \leq N$. Consider the operation $g_j(D) \leftarrow g_j(D)+D^qg_i(D)$. Let the resulting matrix be $G'(D)$ and the associated TBGM be $G_N'^{tb}$. Then $G_N'^{tb}$ is equivalent to $G_N^{tb}$.
\end{pro}
\par
Taking into consideration Proposition 4.3, let us introduce a useful notion. Let $C$ and $C'$ be TB convolutional codes of section length $N$ defined by $G(D)$ and $G'(D)$, respectively. Denote by $L$ and $L'$ the memory lengths of $G(D)$ and $G'(D)$, respectively, where $\max(L, L')+1 \leq N$. Let $G_N^{tb}$ and $G_N'^{tb}$ be the TBGM's associated with $C$ and $C'$, respectively. We see that if $G_N^{tb}$ and $G_N'^{tb}$ are equivalent, then $C=C'$. All of this leads to the following definition.
\begin{df}
When $G_N^{tb}$ and $G_N'^{tb}$ are equivalent, we say that $G(D)$ and $G'(D)$ are ``TB-equivalent''.
\end{df}
Thus we have the following.
\begin{pro}
If $G(D)$ and $G'(D)$ are TB-equivalent, then a TB convolutional code defined by $G(D)$ is represented as a TB convolutional code defined by $G'(D)$, and vice versa.
\end{pro}
\begin{IEEEproof}
A direct consequence of the definition of TB-equivalent.
\end{IEEEproof}


\section{Trellis Reduction for TB Convolutional Codes}
In this section, we will show that for a TB convolutional code of short to moderate section length, the associated TB trellis can be reduced. We begin with an example.

\subsection{An Example of Trellis Reduction}
\begin{figure}[htb]
\begin{center}
\includegraphics[width=8.0cm,clip]{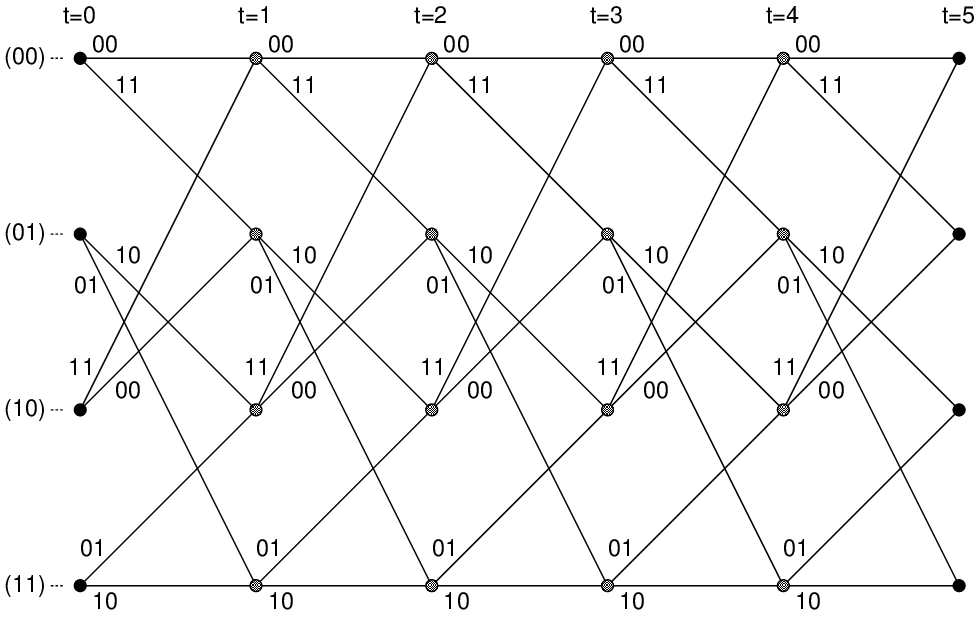}
\end{center}
\caption{The tail-biting convolutional code defined by $G(D)=(1+D+D^2, 1+D^2)$~($N=5$).}
\label{Fig.1}
\end{figure}
Consider the TB convolutional code $C$ defined by $G(D)=(1+D+D^2, 1+D^2)$, where the section length $N$ is set to $5$. The corresponding TB trellis is shown in Fig.1, where the paths which start and end in the same state are TB paths (i.e., valid codewords). Then $C$ is the set of all TB paths. Since $G(D)$ has the polynomial expansion
\begin{eqnarray}
G(D) &= &(1, 1)+(1, 0)D+(1, 1)D^2 \\
&\stackrel{\triangle}{=}& G_0+G_1D+G_2D^2 ,
\end{eqnarray}
the TBGM associated with $C$ is given by
\begin{eqnarray}
G_5^{tb} &=& \left(
\begin{array}{ccccc}
G_0 & G_1 & G_2 & 0 & 0  \\
0 & G_0 & G_1 & G_2 & 0  \\
0 & 0 & G_0 & G_1 & G_2 \\
G_2 & 0 & 0 & G_0 & G_1 \\
G_1 & G_2 & 0 & 0 & G_0
\end{array}
\right) \\
&=& \left(
\begin{array}{cccccccccc}
1 & 1 & 1 & 0 & 1 & 1 & 0 & 0 & 0 & 0 \\
0 & 0 & 1 & 1 & 1 & 0 & 1 & 1 & 0 & 0 \\
0 & 0 & 0 & 0 & 1 & 1 & 1 & 0 & 1 & 1 \\
1 & 1 & 0 & 0 & 0 & 0 & 1 & 1 & 1 & 0 \\
1 & 0 & 1 & 1 & 0 & 0 & 0 & 0 & 1 & 1
\end{array}
\right) .
\end{eqnarray}
Based on $G_5^{tb}$, a characteristic matrix $X$ for $C$ is computed as follows:
\begin{equation}
X=\left(
\begin{array}{cccccccccc}
\mbox{\boldmath $1$} & \mbox{\boldmath $1$} & \mbox{\boldmath $1$} & \mbox{\boldmath $0$} & \mbox{\boldmath $1$} & \mbox{\boldmath $1$} & 0 & 0 & 0 & 0 \\
0 & \mbox{\boldmath $1$} & \mbox{\boldmath $0$}& \mbox{\boldmath $1$} & \mbox{\boldmath $0$} & \mbox{\boldmath $0$} & \mbox{\boldmath $1$} & 0 & 0 & 0 \\
0 & 0 & \mbox{\boldmath $1$} & \mbox{\boldmath $1$} & \mbox{\boldmath $1$} & \mbox{\boldmath $0$} & \mbox{\boldmath $1$} & \mbox{\boldmath $1$} & 0 & 0 \\
0 & 0 & 0 & \mbox{\boldmath $1$} & \mbox{\boldmath $0$}& \mbox{\boldmath $1$} & \mbox{\boldmath $0$} & \mbox{\boldmath $0$} & \mbox{\boldmath $1$} & 0 \\
0 & 0 & 0 & 0 &  \mbox{\boldmath $1$} & \mbox{\boldmath $1$} & \mbox{\boldmath $1$} & \mbox{\boldmath $0$} & \mbox{\boldmath $1$} & \mbox{\boldmath $1$} \\
\mbox{\boldmath $1$} & 0 & 0 & 0 & 0 & \mbox{\boldmath $1$} & \mbox{\boldmath $0$} & \mbox{\boldmath $1$} & \mbox{\boldmath $0$} & \mbox{\boldmath $0$} \\
\mbox{\boldmath $1$} & \mbox{\boldmath $1$} & 0 & 0 & 0 & 0 & \mbox{\boldmath $1$} & \mbox{\boldmath $1$} & \mbox{\boldmath $1$} & \mbox{\boldmath $0$} \\
\mbox{\boldmath $0$} & \mbox{\boldmath $0$} & \mbox{\boldmath $1$} & 0 & 0 & 0 & 0 & \mbox{\boldmath $1$} & \mbox{\boldmath $0$}& \mbox{\boldmath $1$} \\
\mbox{\boldmath $1$} & \mbox{\boldmath $0$} & \mbox{\boldmath $1$} & \mbox{\boldmath $1$} & 0 & 0 & 0 & 0 & \mbox{\boldmath $1$} & \mbox{\boldmath $1$} \\
\mbox{\boldmath $0$}& \mbox{\boldmath $1$} & \mbox{\boldmath $0$} & \mbox{\boldmath $0$} & \mbox{\boldmath $1$} & 0 & 0 & 0 & 0 & \mbox{\boldmath $1$}
\end{array}
\right) 
\begin{array}{l}
$(0, 5]$ \\
$(1, 6]$ \\
$(2, 7]$ \\
$(3, 8]$ \\
$(4, 9]$ \\
$(5, 0]$ \\
$(6, 1]$ \\
$(7, 2]$ \\
$(8, 3]$ \\
$(9, 4]$ .
\end{array}
\end{equation}
Choosing $5$ rows from $X$, let
\begin{equation}
G'=\left(
\begin{array}{cccccccccc}
0 & \mbox{\boldmath $1$} & \mbox{\boldmath $0$}& \mbox{\boldmath $1$} & \mbox{\boldmath $0$} & \mbox{\boldmath $0$} & \mbox{\boldmath $1$} & 0 & 0 & 0 \\
0 & 0 & 0 & \mbox{\boldmath $1$} & \mbox{\boldmath $0$}& \mbox{\boldmath $1$} & \mbox{\boldmath $0$} & \mbox{\boldmath $0$} & \mbox{\boldmath $1$} & 0 \\
\mbox{\boldmath $1$} & 0 & 0 & 0 & 0 & \mbox{\boldmath $1$} & \mbox{\boldmath $0$}& \mbox{\boldmath $1$} & \mbox{\boldmath $0$} & \mbox{\boldmath $0$} \\
\mbox{\boldmath $0$} & \mbox{\boldmath $0$} & \mbox{\boldmath $1$} & 0 & 0 & 0 & 0 & \mbox{\boldmath $1$} & \mbox{\boldmath $0$}& \mbox{\boldmath $1$} \\
\mbox{\boldmath $0$}& \mbox{\boldmath $1$} & \mbox{\boldmath $0$} & \mbox{\boldmath $0$} & \mbox{\boldmath $1$} & 0 & 0 & 0 & 0 & \mbox{\boldmath $1$}
\end{array}
\right)
\begin{array}{l}
$(1, 6]$ \\
$(3, 8]$ \\
$(5, 0]$ \\
$(7, 2]$ \\
$(9, 4]$ .
\end{array}
\end{equation}
We see that the rows of $G'$ are linearly independent and thus generate $C$, i.e., $G'$ is equivalent to $G_5^{tb}$.
\par
Here note that $G'$ consists of the first row and its right cyclic shifts by $i \times 2~(1 \leq i \leq 4)$ positions. Accordingly, $G'$ can be regarded as the TBGM associated with
\begin{eqnarray}
G'(D) &=& (D^3, 1+D) \\
&=& (0, 1)+(0, 1)D+(1, 0)D^3 .
\end{eqnarray}
Hence, $C$ is equally represented as a TB convolutional code defined by $G'(D)=(D^3, 1+D)$. We remark that the constraint length of $G'(D)$ is $\nu'=3$ and is greater than that of $G(D)$.
\par
On the other hand, observe that the first column of $G'(D)$ has a factor $D^2$. Then dividing the first column by $D^2$, we have
\begin{equation}
G'(D)=(D^3, 1+D) \rightarrow G''(D)=(D, 1+D) .
\end{equation}
Note that this transformation corresponds to cyclically shifting the first component of each branch (of a TB path) to the left by two branches (cf. Proposition 4.1). By this transformation, the original TB convolutional code is represented using a trellis associated with $G''(D)$ as well. The trellis for $G''(D)=(D, 1+D)$ is shown in Fig.2. For example, take notice of the TB path in Fig.1 which starts and ends in state $(00)$:
\begin{displaymath}
\mbox{\boldmath $w$}=11~~01~~10~~01~~11 .
\end{displaymath}
Cyclically shifting the first component of each branch to the left by two branches, it becomes
\begin{displaymath}
\mbox{\boldmath $w$}_m=11~~01~~10~~11~~01 .
\end{displaymath}
We see that the modified path is represented as a path which starts and ends in state (1) in Fig.2.
\par
This example shows that there are cases where a given TB convolutional code is represented using a reduced trellis with less state complexity, if we allow partial cyclic shifts of a TB path.
\begin{figure}[htb]
\begin{center}
\includegraphics[width=8.0cm,clip]{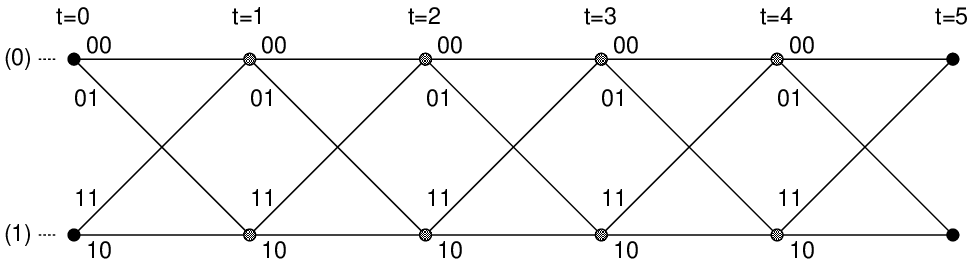}
\end{center}
\caption{The tail-biting convolutional code defined by $G''(D)=(D, 1+D)$~($N=5$).}
\label{Fig.2}
\end{figure}

\subsection{Trellis Reduction for TB Convolutional Codes}
The argument in the previous section, though it was presented in terms of a specific example, is entirely general. Then the method can be directly extended to a general case. Let $G(D)$ be as in Section III. Denote by $\nu$ the constraint length of $G(D)$. Consider a TB convolutional code $C$ of section length $N$ defined by $G(D)$. The trellis reduction procedure becomes as follows.
\par
{\bf Procedure for trellis reduction:}
\begin{itemize}
\item[i)] Compute a characteristic matrix $X$ for $C$ based on the TBGM $G_N^{tb}$, where $X$ consists of $n_0$ rows and their right cyclic shifts by integer multiple of $n_0$.
\item[ii)] Choosing $k$ rows from $X$, form $G'$, where $G'$ has the properties:
\begin{itemize}
\item[1)] The rows of $G'$ are linearly independent and thus generate $C$.
\item[2)] $G'$ consists of $k_0$ rows and their right cyclic shifts by integer multiple of $n_0$.
\end{itemize}
\item[iii)] {\it (Direct reduction)} $G'$ is regarded as the TBGM associated with another generator matrix $G'(D)$. Let $\nu'$ be the constraint length of $G'(D)$. If $\nu'<\nu$, then trellis reduction for $C$ is realized. 
\item[iv)] {\it (Indirect reduction)} Even if $\nu' \geq \nu$, there are cases where $G'(D)$ has a monomial factor $D^p$ in some ($j$th) column. Then there is a possibility that $\nu'$ is reduced by dividing the $j$th column of $G'(D)$ by $D^p$ (the resulting matrix is denoted by $G''(D)$). Let $\nu''$ be the constraint length of $G''(D)$. If $\nu''<\nu$, then the original TB trellis can be reduced. That is, by cyclically shifting the $j$th component of each branch of a TB path ($ \in C$) to the left by $p$ branches, the set of modified paths are equally represented as a TB convolutional code defined by $G''(D)$ (this is justified by Proposition 4.1). Thus trellis reduction is accomplished.
\item[v)] $X$ is not necessarily unique~\cite{glu 111}. Hence, if necessary, try i) $\sim$ iv) using another characteristic matrix $X'$ for $C$.
\end{itemize}
\par
{\it Remark:} For row rate codes, it is rather easy to find $G'$ which is equivalent to $G_N^{tb}$. Also, row rate codes make it easy to determine whether $G'(D)$ can be reduced or not.
\par
As is stated above, there are some restrictions on the selection of $X$ and $G'$. We have the following.
\begin{pro}
The number of characteristic matrices $X \mbox{'}s$ in i) is given by $2^{\theta}$, where $\theta$ is defined in Section III-C. For a fixed $X$, the number of $G'\mbox{'}s$ which satisfy the condition 2) in ii) is given by $_{n_0}\mbox{C}_{k_0}$.
\end{pro}
\begin{IEEEproof}
$G'$ is a candidate for a TBGM associated with an encoder. Hence, the above is a consequence of the structure of TBGM.
\end{IEEEproof}
\par
{\it Example 3:} Consider the TB convolutional code $C$ of section lengh $N=6$ defined by
\begin{equation}
G(D)=(1+D+D^2+D^3, 1+D^2+D^3) .
\end{equation}
Using the associated TBGM, i.e., $G_6^{tb}$, a characteristic matrix $X$ for $C$ is computed as follows:
\begin{equation}
X=\left(
\begin{array}{cccccccccccc}
\mbox{\boldmath $1$} & \mbox{\boldmath $0$}& \mbox{\boldmath $0$} & \mbox{\boldmath $0$} & \mbox{\boldmath $1$} & \mbox{\boldmath $0$} &\mbox{\boldmath $0$} & \mbox{\boldmath $1$} & 0 & 0 & 0 & 0 \\
0 & \mbox{\boldmath $1$} & \mbox{\boldmath $1$} & \mbox{\boldmath $0$} & \mbox{\boldmath $0$} & \mbox{\boldmath $1$} & \mbox{\boldmath $1$} & 0 & 0 & 0 & 0 & 0 \\
0 & 0 & \mbox{\boldmath $1$} & \mbox{\boldmath $0$}& \mbox{\boldmath $0$} & \mbox{\boldmath $0$} & \mbox{\boldmath $1$} & \mbox{\boldmath $0$} &\mbox{\boldmath $0$} & \mbox{\boldmath $1$} & 0 & 0 \\
0 & 0 & 0 & \mbox{\boldmath $1$} & \mbox{\boldmath $1$} & \mbox{\boldmath $0$} & \mbox{\boldmath $0$} & \mbox{\boldmath $1$} & \mbox{\boldmath $1$} & 0 & 0 & 0 \\
0 & 0 & 0 & 0 & \mbox{\boldmath $1$} & \mbox{\boldmath $0$} & \mbox{\boldmath $0$} & \mbox{\boldmath $0$} & \mbox{\boldmath $1$} & \mbox{\boldmath $0$} &\mbox{\boldmath $0$} & \mbox{\boldmath $1$} \\
0 & 0 & 0 & 0 & 0 & \mbox{\boldmath $1$} & \mbox{\boldmath $1$} & \mbox{\boldmath $0$} & \mbox{\boldmath $0$} & \mbox{\boldmath $1$} & \mbox{\boldmath $1$} & 0 \\
\mbox{\boldmath $0$} & \mbox{\boldmath $1$} & 0 & 0 & 0 & 0 & \mbox{\boldmath $1$} & \mbox{\boldmath $0$}& \mbox{\boldmath $0$} & \mbox{\boldmath $0$} & \mbox{\boldmath $1$} & \mbox{\boldmath $0$} \\
\mbox{\boldmath $1$} & 0 & 0 & 0 & 0 & 0 & 0 & \mbox{\boldmath $1$} & \mbox{\boldmath $1$} & \mbox{\boldmath $0$} & \mbox{\boldmath $0$} & \mbox{\boldmath $1$} \\
\mbox{\boldmath $1$} & \mbox{\boldmath $0$} &\mbox{\boldmath $0$} & \mbox{\boldmath $1$} & 0 & 0 & 0 & 0 & \mbox{\boldmath $1$} & \mbox{\boldmath $0$}& \mbox{\boldmath $0$} & \mbox{\boldmath $0$} \\
\mbox{\boldmath $0$} & \mbox{\boldmath $1$} & \mbox{\boldmath $1$} & 0 & 0 & 0 & 0 & 0 & 0 & \mbox{\boldmath $1$} & \mbox{\boldmath $1$} & \mbox{\boldmath $0$} \\
\mbox{\boldmath $0$} & \mbox{\boldmath $0$} & \mbox{\boldmath $1$} & \mbox{\boldmath $0$} &\mbox{\boldmath $0$} & \mbox{\boldmath $1$} & 0 & 0 & 0 & 0 & \mbox{\boldmath $1$} & \mbox{\boldmath $0$} \\
\mbox{\boldmath $1$} & \mbox{\boldmath $0$} & \mbox{\boldmath $0$} & \mbox{\boldmath $1$} & \mbox{\boldmath $1$} & 0 & 0 & 0 & 0 & 0 & 0 & \mbox{\boldmath $1$}
\end{array}
\right)
\begin{array}{l}
$(0, 7]$ \\
$(1, 6]$ \\
$(2, 9]$ \\
$(3, 8]$ \\
$(4, 11]$ \\
$(5, 10]$ \\
$(6, 1]$ \\
$(7, 0]$ \\
$(8, 3]$ \\
$(9, 2]$ \\
$(10, 5]$ \\
$(11, 4]$ .
\end{array}
\end{equation}
Choosing $6$ rows from $X$, define $G'$ as
\begin{equation}
G'=\left(
\begin{array}{cccccccccccc}
\mbox{\boldmath $1$} & \mbox{\boldmath $0$} & \mbox{\boldmath $0$} & \mbox{\boldmath $0$} & \mbox{\boldmath $1$} & \mbox{\boldmath $0$} &\mbox{\boldmath $0$} & \mbox{\boldmath $1$} & 0 & 0 & 0 & 0 \\
0 & 0 & \mbox{\boldmath $1$} & \mbox{\boldmath $0$} & \mbox{\boldmath $0$} & \mbox{\boldmath $0$} & \mbox{\boldmath $1$} & \mbox{\boldmath $0$} &\mbox{\boldmath $0$} & \mbox{\boldmath $1$} & 0 & 0 \\
0 & 0 & 0 & 0 & \mbox{\boldmath $1$} & \mbox{\boldmath $0$} & \mbox{\boldmath $0$} & \mbox{\boldmath $0$} & \mbox{\boldmath $1$} & \mbox{\boldmath $0$} &\mbox{\boldmath $0$} & \mbox{\boldmath $1$} \\
\mbox{\boldmath $0$} & \mbox{\boldmath $1$} & 0 & 0 & 0 & 0 & \mbox{\boldmath $1$} & \mbox{\boldmath $0$}& \mbox{\boldmath $0$} & \mbox{\boldmath $0$} & \mbox{\boldmath $1$} & \mbox{\boldmath $0$} \\
\mbox{\boldmath $1$} & \mbox{\boldmath $0$} &\mbox{\boldmath $0$} & \mbox{\boldmath $1$} & 0 & 0 & 0 & 0 & \mbox{\boldmath $1$} & \mbox{\boldmath $0$}& \mbox{\boldmath $0$} & \mbox{\boldmath $0$} \\
\mbox{\boldmath $0$} & \mbox{\boldmath $0$} & \mbox{\boldmath $1$} & \mbox{\boldmath $0$} &\mbox{\boldmath $0$} & \mbox{\boldmath $1$} & 0 & 0 & 0 & 0 & \mbox{\boldmath $1$} & \mbox{\boldmath $0$}
\end{array}
\right)
\begin{array}{l}
$(0, 7]$ \\
$(2, 9]$ \\
$(4, 11]$ \\
$(6, 1]$ \\
$(8, 3]$ \\
$(10, 5]$ .
\end{array}
\end{equation}
We see that $G'$ is equivalent to $G_6^{tb}$. Also, we see that $G'$ is the TBGM associated with
\begin{equation}
G'(D)=(1+D^2, D^3) .
\end{equation}
Note that the constraint length $\nu'=3$ of $G'(D)$ is not reduced compared to that of $G(D)$. On the other hand, observe that the second column of $G'(D)$ has a factor $D$. Then dividing the column by $D$, we have
\begin{equation}
G'(D)=(1+D^2, D^3) \rightarrow G''(D)=(1+D^2, D^2) .
\end{equation}
This transformation corresponds to cyclically shifting the second component of each branch of a TB path to the left by one branch (cf. Proposition 4.1). As a result, the modified paths are represented using the trellis for $G''(D)$. Thus trellis reduction for $C$ is accomplished.
\par
{\it Remark:} As is stated above, $X$ is not necessarily unique. For example, if a characteristic matrix
\begin{equation}
X'=\left(
\begin{array}{cccccccccccc}
\mbox{\boldmath $1$} & \mbox{\boldmath $1$}& \mbox{\boldmath $1$} & \mbox{\boldmath $0$} & \mbox{\boldmath $1$} & \mbox{\boldmath $1$} &\mbox{\boldmath $1$} & \mbox{\boldmath $1$} & 0 & 0 & 0 & 0 \\
0 & \mbox{\boldmath $1$} & \mbox{\boldmath $1$} & \mbox{\boldmath $0$} & \mbox{\boldmath $0$} & \mbox{\boldmath $1$} & \mbox{\boldmath $1$} & 0 & 0 & 0 & 0 & 0 \\
0 & 0 & \mbox{\boldmath $1$} & \mbox{\boldmath $1$}& \mbox{\boldmath $1$} & \mbox{\boldmath $0$} & \mbox{\boldmath $1$} & \mbox{\boldmath $1$} &\mbox{\boldmath $1$} & \mbox{\boldmath $1$} & 0 & 0 \\
0 & 0 & 0 & \mbox{\boldmath $1$} & \mbox{\boldmath $1$} & \mbox{\boldmath $0$} & \mbox{\boldmath $0$} & \mbox{\boldmath $1$} & \mbox{\boldmath $1$} & 0 & 0 & 0 \\
0 & 0 & 0 & 0 & \mbox{\boldmath $1$} & \mbox{\boldmath $1$} & \mbox{\boldmath $1$} & \mbox{\boldmath $0$} & \mbox{\boldmath $1$} & \mbox{\boldmath $1$} &\mbox{\boldmath $1$} & \mbox{\boldmath $1$} \\
0 & 0 & 0 & 0 & 0 & \mbox{\boldmath $1$} & \mbox{\boldmath $1$} & \mbox{\boldmath $0$} & \mbox{\boldmath $0$} & \mbox{\boldmath $1$} & \mbox{\boldmath $1$} & 0 \\
\mbox{\boldmath $1$} & \mbox{\boldmath $1$} & 0 & 0 & 0 & 0 & \mbox{\boldmath $1$} & \mbox{\boldmath $1$}& \mbox{\boldmath $1$} & \mbox{\boldmath $0$} & \mbox{\boldmath $1$} & \mbox{\boldmath $1$} \\
\mbox{\boldmath $1$} & 0 & 0 & 0 & 0 & 0 & 0 & \mbox{\boldmath $1$} & \mbox{\boldmath $1$} & \mbox{\boldmath $0$} & \mbox{\boldmath $0$} & \mbox{\boldmath $1$} \\
\mbox{\boldmath $1$} & \mbox{\boldmath $1$} &\mbox{\boldmath $1$} & \mbox{\boldmath $1$} & 0 & 0 & 0 & 0 & \mbox{\boldmath $1$} & \mbox{\boldmath $1$}& \mbox{\boldmath $1$} & \mbox{\boldmath $0$} \\
\mbox{\boldmath $0$} & \mbox{\boldmath $1$} & \mbox{\boldmath $1$} & 0 & 0 & 0 & 0 & 0 & 0 & \mbox{\boldmath $1$} & \mbox{\boldmath $1$} & \mbox{\boldmath $0$} \\
\mbox{\boldmath $1$} & \mbox{\boldmath $0$} & \mbox{\boldmath $1$} & \mbox{\boldmath $1$} &\mbox{\boldmath $1$} & \mbox{\boldmath $1$} & 0 & 0 & 0 & 0 & \mbox{\boldmath $1$} & \mbox{\boldmath $1$} \\
\mbox{\boldmath $1$} & \mbox{\boldmath $0$} & \mbox{\boldmath $0$} & \mbox{\boldmath $1$} & \mbox{\boldmath $1$} & 0 & 0 & 0 & 0 & 0 & 0 & \mbox{\boldmath $1$}
\end{array}
\right)
\begin{array}{l}
$(0, 7]$ \\
$(1, 6]$ \\
$(2, 9]$ \\
$(3, 8]$ \\
$(4, 11]$ \\
$(5, 10]$ \\
$(6, 1]$ \\
$(7, 0]$ \\
$(8, 3]$ \\
$(9, 2]$ \\
$(10, 5]$ \\
$(11, 4]$
\end{array}
\end{equation}
is used, then trellis reduction cannot be realized using the above procedure.
\par
Using appropriate characteristic matrices, the above reduction method can also be applied to the following cases:
\begin{itemize}
\item[(1)] $R=1/2,~\nu=4,~N=6:$
\begin{eqnarray}
G(D) &=& (1+D+D^4, 1+D^2+D^3+D^4) \nonumber \\
&\rightarrow& G'(D)=(D^2, 1+D) \nonumber \\
&\rightarrow& G''(D)=(D, 1+D) . \nonumber
\end{eqnarray}
\item[(2)] $R=1/2,~\nu=5,~N=10:$
\begin{eqnarray}
G(D) &=& (1+D+D^2+D^3+D^4+D^5, 1+D^3+D^5) \nonumber \\
&\rightarrow& G'(D)=(D^4+D^5, 1+D+D^4) \nonumber \\
&\rightarrow& G''(D)=(D^3+D^4, 1+D+D^4) . \nonumber
\end{eqnarray}
\item[(3)] $R=1/2,~\nu=6,~N=8:$
\begin{eqnarray}
G(D) &=& (1+D+D^4+D^5+D^6, 1+D^2+D^3+D^4+D^6) \nonumber \\
&\rightarrow& G'(D)=(D+D^2+D^3, 1+D^2+D^3) . \nonumber
\end{eqnarray}
\item[(4)] $R=1/2,~\nu=6,~N=8:$
\begin{eqnarray}
G(D) &=& (1+D+D^2+D^3+D^6, 1+D^2+D^3+D^5+D^6) \nonumber \\
&\rightarrow& G'(D)=(1+D^2+D^3, D^2+D^3+D^4) \nonumber \\
&\rightarrow& G''(D)=(1+D^2+D^3, D+D^2+D^3) . \nonumber
\end{eqnarray}
\item[(5)] $R=1/3,~\nu=3,~N=5:$
\begin{eqnarray}
G(D) &=& (1+D+D^2+D^3, 1+D+D^3, 1+D^2+D^3) \nonumber \\
&\rightarrow& G'(D)=(D+D^2, D^3, 1) \nonumber \\
&\rightarrow& G''(D)=(1+D, D, 1) . \nonumber
\end{eqnarray}
\item[(6)] $R=2/3,~\nu=4,~N=6:$
\begin{eqnarray}
G(D) &=& \left(
\begin{array}{ccc}
D+D^2 & 1+D & 1+D+D^2 \\
1 & 1+D^2 & 1+D^2
\end{array}
\right) \nonumber \\
&\rightarrow& G'(D)=\left(
\begin{array}{ccc}
D^2 & 1+D & D \\
D & 0 & 1+D^2
\end{array}
\right) \nonumber \\
&\rightarrow& G''(D)=\left(
\begin{array}{ccc}
D & 1+D & D \\
1 & 0 & 1+D^2
\end{array}
\right) . \nonumber
\end{eqnarray}
\end{itemize}

\subsection{Trellis Reduction Using a Reciprocal Dual Encoder}
For high rate codes, $G'(D)$ may not have a monomial factor in any columns. Then it is not easily determined whether $G'(D)$ can be reduced or not. In such cases, it is useful to consider a reciprocal dual encoder $\tilde H'(D)$ associated with $G'(D)$. A reciprocal dual encoder~\cite{ried 98} is defined as follows: Let $G(D)$ be as in Section III. Also, let $H(D)$ be a corresponding check matrix with size $(n_0-k_0) \times n_0$. A reciprocal dual encoder $\tilde H(D)$ is obtained by substituting $D^{-1}$ for $D$ in $H(D)$ and by multiplying the $i$th ($1 \leq i \leq n_0-k_0$) row of the resulting matrix by $D^{\nu_i^{\perp}}$, where $\nu_i^{\perp}$ is the degree of the $i$th row of $H(D)$.
\begin{df}[McEliece and Lin~\cite{mc 962}]
Let $G_{scalar}$ be a scalar generator matrix for a {\it terminated} convolutional code defined by $G(D)$~\cite{mc 962,ried 98}. $G_{scalar}$ is given by
\begin{equation}
\arraycolsep=1pt
G_{scalar}=\left(
\arraycolsep=1pt
\begin{array}{cccccccccc}
G_0 & G_1 & \scriptstyle{\ldots} & G_L &  &  &  &  &  &  \\
 & G_0 & \scriptstyle{\ldots} & G_{L-1} & G_L &  &  &  &  &  \\
 &  & \scriptstyle{\ldots} & \scriptstyle{\ldots} & \scriptstyle{\ldots} & \scriptstyle{\ldots} &  &  &  &  \\
 &  &  & G_0 & G_1 & \scriptstyle{\ldots} & G_L &  &  &  \\
 &  &  &  & G_0 & \scriptstyle{\ldots} & G_{L-1} & G_L &  &  \\
 &  &  &  &  & \scriptstyle{\ldots} & \scriptstyle{\ldots} & \scriptstyle{\ldots} & \scriptstyle{\ldots} &  \\
 &  &  &  &  &  & G_0 & G_1 & \scriptstyle{\ldots} & G_L
\end{array}
\right) .
\end{equation}
The $(L+1)k_0 \times n_0$ matrix
\begin{equation}
\hat G\stackrel{\triangle}{=}\left(
\begin{array}{c}
G_L \\
G_{L-1} \\
\cdots \\
G_0
\end{array}
\right) ,
\end{equation}
which repeatedly appears as a vertical slice in $G_{scalar}$ except initial and final transient sections, is called the {\it matrix module}. Then the {\it trellis module} $\mathcal{T}_{0}$ for the trellis associated with $G_{scalar}$ corresponds to $\hat G$. If $G_{scalar}$ is in minimal-span form, then $\mathcal{T}_{0}$ is {\it minimal}. The {\it state complexity profile} of $\mathcal{T}_{0}$ is an $n_0$-tuple consisting of the dimensions of state spaces $V_i~(0 \leq i \leq n_0-1)$ of $\mathcal{T}_{0}$.
\end{df}
\par
The meaning of obtaining a reciprocal dual encoder is based on the following result~\cite{taji 04,tan 02,tan 06}.
\begin{pro}[Tang and Lin~\cite{tan 02}]
Consider a minimal trellis module of $G(D)$ and that of an associated reciprocal dual encoder $\tilde H(D)$. Then their state complex profiles are identical.
\end{pro}
\par
Hence, in order to determine whether $G'(D)$ is reduced or not, we can compute a reciprocal dual encoder $\tilde H'(D)$ associated with $G'(D)$. In connection with an encoder $G(D)$ and an associated reciprocal dual encoder $\tilde H(D)$, we have the following.
\begin{pro}
Let $G_N^{tb}$ be the TBGM associated with $G(D)$. Then a check matrix corresponding to $G_N^{tb}$ is obtained as the TBGM (denoted by $\tilde H_N^{tb}$) associated with a reciprocal dual encoder $\tilde H(D)$.
\end{pro}
\begin{IEEEproof}
Let the polynomial expansion of $H(D)$ be
\begin{equation}
H(D)=H_0+H_1D+\cdots +H_{M-1}D^{M-1}+H_MD^M ,
\end{equation}
where $M$ is the memory length of $H(D)$ and $H_i~(0 \leq i \leq M)$ are $(n_0-k_0) \times n_0$ matrices. It is known (e.g.,~\cite{taji 122}) that a check matrix corresponding to $G_N^{tb}$ is given by
\begin{equation}
\arraycolsep=1pt
H^{tb}=\left(
\arraycolsep=1pt
\begin{array}{cccccccc}
H_0 &  &  &  & H_M & \scriptstyle{\ldots} & H_2 & H_1 \\
H_1 & H_0 &  &  &  & \scriptstyle{\ldots} & \scriptstyle{\ldots} & H_2 \\
\scriptstyle{\ldots} & H_1 & \scriptstyle{\ldots} &  &  &  & H_M & \scriptstyle{\ldots} \\
H_{M-1} & \scriptstyle{\ldots} & \scriptstyle{\ldots} & H_0 &  &  &  & H_M \\
H_M & H_{M-1} & \scriptstyle{\ldots} & H_1 & H_0 &  &  &  \\
 & H_M & \scriptstyle{\ldots} & \scriptstyle{\ldots} & H_1 & \scriptstyle{\ldots} &  &  \\
 &  & \scriptstyle{\ldots} & H_{M-1} & \scriptstyle{\ldots} & \scriptstyle{\ldots} & H_0 &  \\
 &  &  & H_M & H_{M-1} & \scriptstyle{\ldots} & H_1 & H_0
\end{array}
\right)
\end{equation}
with size $(n_0-k_0)N \times n_0N$.
On the other hand, let the polynomial expansion of $\tilde H(D)$ be
\begin{equation}
\tilde H(D)=\tilde H_0+\tilde H_1D+\cdots +\tilde H_{M-1}D^{M-1}+\tilde H_MD^M .
\end{equation}
Then the TBGM associated with $\tilde H(D)$ (denoted by $\tilde H_N^{tb}$) is defined by
\begin{equation}
\arraycolsep=1pt
\tilde H_N^{tb}=\left(
\arraycolsep=1pt
\begin{array}{cccccccc}
\tilde H_0 & \tilde H_1 & \scriptstyle{\ldots} & \tilde H_{M-1} & \tilde H_M &  &  &  \\
 & \tilde H_0 & \scriptstyle{\ldots} & \scriptstyle{\ldots} & \tilde H_{M-1} & \tilde H_M &  &  \\
 &  & \scriptstyle{\ldots} & \scriptstyle{\ldots} & \scriptstyle{\ldots} & \scriptstyle{\ldots} & \scriptstyle{\ldots} &  \\
 &  &  & \tilde H_0 & \tilde H_1 & \scriptstyle{\ldots} & \scriptstyle{\ldots} & \tilde H_M \\
\tilde H_M &  &  &  & \tilde H_0 & \tilde H_1 & \scriptstyle{\ldots} & \tilde H_{M-1} \\
\tilde H_{M-1} & \tilde H_M &  &  &  & \tilde H_0 & \scriptstyle{\ldots} & \scriptstyle{\ldots} \\
\scriptstyle{\ldots} & \scriptstyle{\ldots} & \scriptstyle{\ldots} &  &  &  & \scriptstyle{\ldots} & \tilde H_1 \\
\tilde H_1 & \tilde H_2 & \scriptstyle{\ldots} & \tilde H_M &  &  &  & \tilde H_0
\end{array}
\right) 
\end{equation}
with size $(n_0-k_0)N \times n_0N$.
\par
Here take notice of the $i$th ($1 \leq i \leq n_0-k_0$) row of
\begin{displaymath}
\left(\tilde H_0, \tilde H_1, \cdots, \tilde H_{M-1}, \tilde H_M, 0, \cdots, 0 \right)\subset \tilde H_N^{tb} .
\end{displaymath}
We see that the row is identical to the $i$th row of
\begin{displaymath}
\left(H_{\nu_i^{\perp}}, \cdots, H_1, H_0, 0, \cdots, 0, H_M, \cdots, H_{\nu_i^{\perp}+1} \right)\subset H^{tb} .
\end{displaymath}
Similarly, the $i$th ($1 \leq i \leq n_0-k_0$) row of
\begin{displaymath}
\left(0, \tilde H_0, \tilde H_1, \cdots, \tilde H_{M-1}, \tilde H_M, 0, \cdots, 0 \right)\subset \tilde H_N^{tb}
\end{displaymath}
is identical to the $i$th row of
\begin{displaymath}
\left(H_{\nu_i^{\perp}+1}, \cdots, H_1, H_0, 0, \cdots, 0, H_M, \cdots, H_{\nu_i^{\perp}+2} \right)\subset H^{tb} .
\end{displaymath}
Due to the cyclic structures of $H^{tb}$ and $\tilde H_N^{tb}$, similar correspondences hold successively. Hence, $\tilde H_N^{tb}$ is given as a row permutation of $H^{tb}$.
\end{IEEEproof}
\par
A procedure for computing $\tilde H'(D)$ is obtained based on the above proposition.
\par
{\bf Procedure for computing $\tilde H'(D)$:}
\begin{itemize}
\item[i)] Compute a characteristic matrix $Y$ for the dual code $C^{\perp}$ based on $\tilde H_N^{tb}$, where $Y$ consists of $n_0$ rows and their right cyclic shifts by integer multiple of $n_0$.
\item[ii)] Choosing $(n-k)$ rows from $Y$, form $\tilde H'$, where $\tilde H'$ has the properties:
\begin{itemize}
\item[1)] The rows of $\tilde H'$ are linearly independent and thus generate $C^{\perp}$.
\item[2)] $\tilde H'$ consists of $(n_0-k_0)$ rows and their right cyclic shifts by integer multiple of $n_0$.
\end{itemize}
\item[iii)] Let $\tilde H'(D)$ be the polynomial matrix whose TBGM is $\tilde H'$.
\item[iv)] Note that $G'$ and $\tilde H'$ are equivalent to $G_N^{tb}$ and $\tilde H_N^{tb}$, respectively. Hence, $\tilde H'$ is a check matrix corresponding to $G'$. Then it follows from Proposition 5.3 that $\tilde H'(D)$ is a reciprocal dual encoder associated with $G'(D)$.
\item[v)] $Y$ is not necessarily unique. Hence, if necessary, try i) $\sim$ iv) using another characteristic matrix $Y'$ for $C^{\perp}$.
\end{itemize}
\par
The following is an example where trellis reduction is realized using a reciprocal dual encoder.
\par
{\it Example 4:} Consider the rate $R=2/3$ TB convolutional code $C$ of section length $N=5$ with generator matrix
\begin{eqnarray}
G(D) &=& \left(
\begin{array}{ccc}
1+D & D & 1 \\
D^2 & 1 & 1+D+D^2
\end{array}
\right) \\
&=& \left(
\begin{array}{ccc}
1 & 0 & 1 \\
0 & 1 & 1
\end{array}
\right)+\left(
\begin{array}{ccc}
1 & 1 & 0 \\
0 & 0 & 1
\end{array}
\right)D+\left(
\begin{array}{ccc}
0 & 0 & 0 \\
1 & 0 & 1
\end{array}
\right)D^2 \nonumber \\
&\stackrel{\triangle}{=}& G_0+G_1D+G_2D^2 .
\end{eqnarray}
Based on the associated TBGM, i.e., 
\begin{equation}
G_5^{tb}=\left(
\begin{array}{ccccc}
G_0 & G_1 & G_2 & 0 & 0  \\
0 & G_0 & G_1 & G_2 & 0  \\
0 & 0 & G_0 & G_1 & G_2 \\
G_2 & 0 & 0 & G_0 & G_1 \\
G_1 & G_2 & 0 & 0 & G_0
\end{array}
\right) ,
\end{equation}
a characteristic matrix $X$ for $C$ is computed as follows:
\begin{equation}
X=\left(
\begin{array}{ccccccccccccccc}
\mbox{\boldmath $1$} & \mbox{\boldmath $0$} & \mbox{\boldmath $1$} & \mbox{\boldmath $1$} & \mbox{\boldmath $1$} & 0 & 0 & 0 & 0 & 0 & 0 & 0 & 0 & 0 & 0 \\
0 & \mbox{\boldmath $1$} & \mbox{\boldmath $1$}& \mbox{\boldmath $0$} & \mbox{\boldmath $0$} & \mbox{\boldmath $1$} & \mbox{\boldmath $1$} & \mbox{\boldmath $0$} & \mbox{\boldmath $1$} & 0 & 0 & 0 & 0 & 0 & 0 \\
0 & 0 & \mbox{\boldmath $1$} & \mbox{\boldmath $0$} & \mbox{\boldmath $0$} & \mbox{\boldmath $1$} & \mbox{\boldmath $1$} & 0 & 0 & 0 & 0 & 0 & 0 & 0 & 0 \\
0 & 0 & 0 & \mbox{\boldmath $1$} & \mbox{\boldmath $0$} & \mbox{\boldmath $1$} & \mbox{\boldmath $1$} & \mbox{\boldmath $1$} & 0 & 0 & 0 & 0 & 0 & 0 & 0 \\
0 & 0 & 0 & 0 & \mbox{\boldmath $1$} & \mbox{\boldmath $1$}& \mbox{\boldmath $0$} & \mbox{\boldmath $0$} & \mbox{\boldmath $1$} & \mbox{\boldmath $1$} & \mbox{\boldmath $0$} & \mbox{\boldmath $1$} & 0 & 0 & 0 \\
0 & 0 & 0 & 0 & 0 & \mbox{\boldmath $1$} & \mbox{\boldmath $0$} & \mbox{\boldmath $0$} & \mbox{\boldmath $1$} & \mbox{\boldmath $1$} & 0 & 0 & 0 & 0 & 0 \\
0 & 0 & 0 & 0 & 0 & 0 & \mbox{\boldmath $1$} & \mbox{\boldmath $0$} & \mbox{\boldmath $1$} & \mbox{\boldmath $1$} & \mbox{\boldmath $1$} & 0 & 0 & 0 & 0 \\
0 & 0 & 0 & 0 & 0 & 0 & 0 & \mbox{\boldmath $1$} & \mbox{\boldmath $1$}& \mbox{\boldmath $0$} & \mbox{\boldmath $0$} & \mbox{\boldmath $1$} & \mbox{\boldmath $1$} & \mbox{\boldmath $0$} & \mbox{\boldmath $1$} \\
0 & 0 & 0 & 0 & 0 & 0 & 0 & 0 & \mbox{\boldmath $1$} & \mbox{\boldmath $0$} & \mbox{\boldmath $0$} & \mbox{\boldmath $1$} & \mbox{\boldmath $1$} & 0 & 0 \\
0 & 0 & 0 & 0 & 0 & 0 & 0 & 0 & 0 & \mbox{\boldmath $1$} & \mbox{\boldmath $0$} & \mbox{\boldmath $1$} & \mbox{\boldmath $1$} & \mbox{\boldmath $1$} & 0 \\
\mbox{\boldmath $1$} & \mbox{\boldmath $0$} & \mbox{\boldmath $1$} & 0 & 0 & 0 & 0 & 0 & 0 & 0 & \mbox{\boldmath $1$} & \mbox{\boldmath $1$}& \mbox{\boldmath $0$} & \mbox{\boldmath $0$} & \mbox{\boldmath $1$} \\
\mbox{\boldmath $1$} & 0 & 0 & 0 & 0 & 0 & 0 & 0 & 0 & 0 & 0 & \mbox{\boldmath $1$} & \mbox{\boldmath $0$} & \mbox{\boldmath $0$} & \mbox{\boldmath $1$} \\
\mbox{\boldmath $1$} & \mbox{\boldmath $1$} & 0 & 0 & 0 & 0 & 0 & 0 & 0 & 0 & 0 & 0 & \mbox{\boldmath $1$} & \mbox{\boldmath $0$} & \mbox{\boldmath $1$} \\
\mbox{\boldmath $0$} & \mbox{\boldmath $0$} & \mbox{\boldmath $1$} & \mbox{\boldmath $1$} & \mbox{\boldmath $0$} & \mbox{\boldmath $1$} & 0 & 0 & 0 & 0 & 0 & 0 & 0 & \mbox{\boldmath $1$} & \mbox{\boldmath $1$} \\
\mbox{\boldmath $0$} & \mbox{\boldmath $0$} & \mbox{\boldmath $1$} & \mbox{\boldmath $1$} & 0 & 0 & 0 & 0 & 0 & 0 & 0 & 0 & 0 & 0 & \mbox{\boldmath $1$}
\end{array}
\right) .
\end{equation}
The span list for $X$ is given by
\begin{eqnarray}
T &=& \{(0, 4], (1, 8], (2, 6], (3, 7], (4, 11], (5, 9], (6, 10], (7, 14], \nonumber \\
&& (8, 12], (9, 13], (10, 2], (11, 0], (12, 1], (13, 5], (14, 3]\} .
\end{eqnarray}
Choosing $10$ rows from $X$, let
\begin{equation}
G'=\left(
\begin{array}{ccccccccccccccc}
\mbox{\boldmath $1$} & \mbox{\boldmath $0$} & \mbox{\boldmath $1$} & \mbox{\boldmath $1$} & \mbox{\boldmath $1$} & 0 & 0 & 0 & 0 & 0 & 0 & 0 & 0 & 0 & 0 \\
0 & 0 & \mbox{\boldmath $1$} & \mbox{\boldmath $0$} & \mbox{\boldmath $0$} & \mbox{\boldmath $1$} & \mbox{\boldmath $1$} & 0 & 0 & 0 & 0 & 0 & 0 & 0 & 0 \\
0 & 0 & 0 & \mbox{\boldmath $1$} & \mbox{\boldmath $0$} & \mbox{\boldmath $1$} & \mbox{\boldmath $1$} & \mbox{\boldmath $1$} & 0 & 0 & 0 & 0 & 0 & 0 & 0 \\
0 & 0 & 0 & 0 & 0 & \mbox{\boldmath $1$} & \mbox{\boldmath $0$} & \mbox{\boldmath $0$} & \mbox{\boldmath $1$} & \mbox{\boldmath $1$} & 0 & 0 & 0 & 0 & 0 \\
0 & 0 & 0 & 0 & 0 & 0 & \mbox{\boldmath $1$} & \mbox{\boldmath $0$} & \mbox{\boldmath $1$} & \mbox{\boldmath $1$} & \mbox{\boldmath $1$} & 0 & 0 & 0 & 0 \\
0 & 0 & 0 & 0 & 0 & 0 & 0 & 0 & \mbox{\boldmath $1$} & \mbox{\boldmath $0$} & \mbox{\boldmath $0$} & \mbox{\boldmath $1$} & \mbox{\boldmath $1$} & 0 & 0 \\
0 & 0 & 0 & 0 & 0 & 0 & 0 & 0 & 0 & \mbox{\boldmath $1$} & \mbox{\boldmath $0$} & \mbox{\boldmath $1$} & \mbox{\boldmath $1$} & \mbox{\boldmath $1$} & 0 \\
\mbox{\boldmath $1$} & 0 & 0 & 0 & 0 & 0 & 0 & 0 & 0 & 0 & 0 & \mbox{\boldmath $1$} & \mbox{\boldmath $0$} & \mbox{\boldmath $0$} & \mbox{\boldmath $1$} \\
\mbox{\boldmath $1$} & \mbox{\boldmath $1$} & 0 & 0 & 0 & 0 & 0 & 0 & 0 & 0 & 0 & 0 & \mbox{\boldmath $1$} & \mbox{\boldmath $0$} & \mbox{\boldmath $1$} \\
\mbox{\boldmath $0$} & \mbox{\boldmath $0$} & \mbox{\boldmath $1$} & \mbox{\boldmath $1$} & 0 & 0 & 0 & 0 & 0 & 0 & 0 & 0 & 0 & 0 & \mbox{\boldmath $1$}
\end{array}
\right) .
\end{equation}
The span list for $G'$ is given by
\begin{eqnarray}
S &=& \{(0, 4], (2, 6], (3, 7], (5, 9], (6, 10], \nonumber \\
&& (8, 12], (9, 13], (11, 0], (12, 1], (14, 3]\} .
\end{eqnarray}
We see that the rows of $G'$ are linearly independent and thus generate $C$, i.e., $G'$ is equivalent to $G_5^{tb}$. Also, note that $G'$ is the TBGM associated with
\begin{eqnarray}
G'(D) &=& \left(
\begin{array}{ccc}
1+D & D & 1 \\
D^2 & 0 & 1+D
\end{array}
\right) \\
&=& \left(
\begin{array}{ccc}
1 & 0 & 1 \\
0 & 0 & 1
\end{array}
\right)+\left(
\begin{array}{ccc}
1 & 1 & 0 \\
0 & 0 & 1
\end{array}
\right)D+\left(
\begin{array}{ccc}
0 & 0 & 0 \\
1 & 0 & 0
\end{array}
\right)D^2 \nonumber \\
&\stackrel{\triangle}{=}& G_0'+G_1'D+G_2'D^2 .
\end{eqnarray}
Hence, the original TB convolutional code is equally represented as a TB convolutional code defined by $G'(D)$.
\par
Observe that the constraint length of $G'(D)$ is $\nu'=3$ and is equal to that of $G(D)$. Also, notice that the second column of $G'(D)$ has a factor $D$. However, $\nu'$ is not reduced by dividing the column by $D$. In general, it is difficult to tell a possibility of reduction of $G'(D)$ just by looking at its entries. So, we will compute a reciprocal dual encoder $\tilde H'(D)$ associated with $G'(D)$.
\par
We begin with a reciprocal dual encoder $\tilde H(D)$ associated with $G(D)$. $\tilde H(D)$ is given by
\begin{equation}
\tilde H(D)=(1+D+D^2+D^3, 1+D+D^3, 1+D^2+D^3) .
\end{equation}
Based on $\tilde H_5^{tb}$, a characteristic matrix $Y$ for $C^{\perp}$ is computed as follows:
\begin{equation}
Y=\left(
\begin{array}{ccccccccccccccc}
\mbox{\boldmath $1$} & \mbox{\boldmath $1$} & \mbox{\boldmath $1$} & \mbox{\boldmath $1$} & \mbox{\boldmath $1$} & \mbox{\boldmath $0$} & \mbox{\boldmath $1$} & \mbox{\boldmath $0$} & \mbox{\boldmath $1$} & \mbox{\boldmath $1$} & \mbox{\boldmath $1$} & \mbox{\boldmath $1$} & 0 & 0 & 0 \\
0 & \mbox{\boldmath $1$} & \mbox{\boldmath $0$} & \mbox{\boldmath $0$} & \mbox{\boldmath $0$} & \mbox{\boldmath $0$} & \mbox{\boldmath $0$} & \mbox{\boldmath $0$} & \mbox{\boldmath $1$} & \mbox{\boldmath $1$} & \mbox{\boldmath $0$} & \mbox{\boldmath $0$} & \mbox{\boldmath $1$} & 0 & 0  \\
0 & 0 & \mbox{\boldmath $1$} & \mbox{\boldmath $1$} & \mbox{\boldmath $0$} & \mbox{\boldmath $0$} & \mbox{\boldmath $1$} & \mbox{\boldmath $0$} & \mbox{\boldmath $0$} & \mbox{\boldmath $0$} & \mbox{\boldmath $1$} & 0 & 0 & 0 & 0 \\
0 & 0 & 0 & \mbox{\boldmath $1$} & \mbox{\boldmath $1$} & \mbox{\boldmath $1$} & \mbox{\boldmath $1$} & \mbox{\boldmath $1$} & \mbox{\boldmath $0$} & \mbox{\boldmath $1$} & \mbox{\boldmath $0$} & \mbox{\boldmath $1$} & \mbox{\boldmath $1$} & \mbox{\boldmath $1$} & \mbox{\boldmath $1$} \\
\mbox{\boldmath $1$} & 0 & 0 & 0 & \mbox{\boldmath $1$} & \mbox{\boldmath $0$} & \mbox{\boldmath $0$} & \mbox{\boldmath $0$} & \mbox{\boldmath $0$} & \mbox{\boldmath $0$} & \mbox{\boldmath $0$} & \mbox{\boldmath $1$} & \mbox{\boldmath $1$} & \mbox{\boldmath $0$} & \mbox{\boldmath $0$} \\
0 & 0 & 0 & 0 & 0 & \mbox{\boldmath $1$} & \mbox{\boldmath $1$} & \mbox{\boldmath $0$} & \mbox{\boldmath $0$} & \mbox{\boldmath $1$} & \mbox{\boldmath $0$} & \mbox{\boldmath $0$} & \mbox{\boldmath $0$} & \mbox{\boldmath $1$} & 0 \\
\mbox{\boldmath $1$} & \mbox{\boldmath $1$} & \mbox{\boldmath $1$} & 0 & 0 & 0 & \mbox{\boldmath $1$} & \mbox{\boldmath $1$} & \mbox{\boldmath $1$} & \mbox{\boldmath $1$} & \mbox{\boldmath $1$} & \mbox{\boldmath $0$} & \mbox{\boldmath $1$} & \mbox{\boldmath $0$} & \mbox{\boldmath $1$} \\
\mbox{\boldmath $1$} & \mbox{\boldmath $0$} & \mbox{\boldmath $0$} & \mbox{\boldmath $1$} & 0 & 0 & 0 & \mbox{\boldmath $1$} & \mbox{\boldmath $0$} & \mbox{\boldmath $0$} & \mbox{\boldmath $0$} & \mbox{\boldmath $0$} & \mbox{\boldmath $0$} & \mbox{\boldmath $0$} & \mbox{\boldmath $1$} \\
\mbox{\boldmath $0$} & \mbox{\boldmath $1$} & 0 & 0 & 0 & 0 & 0 & 0 & \mbox{\boldmath $1$} & \mbox{\boldmath $1$} & \mbox{\boldmath $0$} & \mbox{\boldmath $0$} & \mbox{\boldmath $1$} & \mbox{\boldmath $0$} & \mbox{\boldmath $0$} \\
\mbox{\boldmath $1$} & \mbox{\boldmath $0$} & \mbox{\boldmath $1$} & \mbox{\boldmath $1$} & \mbox{\boldmath $1$} & \mbox{\boldmath $1$} & 0 & 0 & 0 & \mbox{\boldmath $1$} & \mbox{\boldmath $1$} & \mbox{\boldmath $1$} & \mbox{\boldmath $1$} & \mbox{\boldmath $1$} & \mbox{\boldmath $0$} \\
\mbox{\boldmath $0$} & \mbox{\boldmath $0$} & \mbox{\boldmath $1$} & \mbox{\boldmath $1$} & \mbox{\boldmath $0$} & \mbox{\boldmath $0$} & \mbox{\boldmath $1$} & 0 & 0 & 0 & \mbox{\boldmath $1$} & \mbox{\boldmath $0$} & \mbox{\boldmath $0$} & \mbox{\boldmath $0$} & \mbox{\boldmath $0$} \\
\mbox{\boldmath $1$} & \mbox{\boldmath $0$} & \mbox{\boldmath $0$} & \mbox{\boldmath $0$} & \mbox{\boldmath $1$} & 0 & 0 & 0 & 0 & 0 & 0 & \mbox{\boldmath $1$} & \mbox{\boldmath $1$} & \mbox{\boldmath $0$} & \mbox{\boldmath $0$} \\
\mbox{\boldmath $1$} & \mbox{\boldmath $1$} & \mbox{\boldmath $0$} & \mbox{\boldmath $1$} & \mbox{\boldmath $0$} & \mbox{\boldmath $1$} & \mbox{\boldmath $1$} & \mbox{\boldmath $1$} & \mbox{\boldmath $1$} & 0 & 0 & 0 & \mbox{\boldmath $1$} & \mbox{\boldmath $1$} & \mbox{\boldmath $1$} \\
\mbox{\boldmath $0$} & \mbox{\boldmath $0$} & \mbox{\boldmath $0$} & \mbox{\boldmath $0$} & \mbox{\boldmath $0$} & \mbox{\boldmath $1$} & \mbox{\boldmath $1$} & \mbox{\boldmath $0$} & \mbox{\boldmath $0$} & \mbox{\boldmath $1$} & 0 & 0 & 0 & \mbox{\boldmath $1$} & \mbox{\boldmath $0$} \\
\mbox{\boldmath $1$} & \mbox{\boldmath $0$} & \mbox{\boldmath $0$} & \mbox{\boldmath $1$} & \mbox{\boldmath $0$} & \mbox{\boldmath $0$} & \mbox{\boldmath $0$} & \mbox{\boldmath $1$} & 0 & 0 & 0 & 0 & 0 & 0 & \mbox{\boldmath $1$}
\end{array}
\right) .
\end{equation}
The span list for $Y$ is given by
\begin{eqnarray}
\hat T &=& \{(0, 11], (1, 12], (2, 10], (3, 14], (4, 0], (5, 13], (6, 2], (7, 3], \nonumber \\
&& (8, 1], (9, 5], (10, 6], (11, 4], (12, 8], (13, 9], (14, 7]\} .
\end{eqnarray}
Note that if the span list for $X$ is $T=\{(a_l, b_l],~l=1, \cdots , 15\}$, then the span list for $Y$ is given by $\hat T=\{(b_l, a_l],~l=1, \cdots , 15\}$~\cite{glu 112,koe 03}.
\par
Next, choosing $5$ rows from $Y$, let
\begin{equation}
\tilde H'=\left(
\begin{array}{ccccccccccccccc}
0 & 0 & \mbox{\boldmath $1$} & \mbox{\boldmath $1$} & \mbox{\boldmath $0$} & \mbox{\boldmath $0$} & \mbox{\boldmath $1$} & \mbox{\boldmath $0$} & \mbox{\boldmath $0$} & \mbox{\boldmath $0$} & \mbox{\boldmath $1$} & 0 & 0 & 0 & 0 \\
0 & 0 & 0 & 0 & 0 & \mbox{\boldmath $1$} & \mbox{\boldmath $1$} & \mbox{\boldmath $0$} & \mbox{\boldmath $0$} & \mbox{\boldmath $1$} & \mbox{\boldmath $0$} & \mbox{\boldmath $0$} & \mbox{\boldmath $0$} & \mbox{\boldmath $1$} & 0 \\
\mbox{\boldmath $0$} & \mbox{\boldmath $1$} & 0 & 0 & 0 & 0 & 0 & 0 & \mbox{\boldmath $1$} & \mbox{\boldmath $1$} & \mbox{\boldmath $0$} & \mbox{\boldmath $0$} & \mbox{\boldmath $1$} & \mbox{\boldmath $0$} & \mbox{\boldmath $0$} \\
\mbox{\boldmath $1$} & \mbox{\boldmath $0$} & \mbox{\boldmath $0$} & \mbox{\boldmath $0$} & \mbox{\boldmath $1$} & 0 & 0 & 0 & 0 & 0 & 0 & \mbox{\boldmath $1$} & \mbox{\boldmath $1$} & \mbox{\boldmath $0$} & \mbox{\boldmath $0$} \\
\mbox{\boldmath $1$} & \mbox{\boldmath $0$} & \mbox{\boldmath $0$} & \mbox{\boldmath $1$} & \mbox{\boldmath $0$} & \mbox{\boldmath $0$} & \mbox{\boldmath $0$} & \mbox{\boldmath $1$} & 0 & 0 & 0 & 0 & 0 & 0 & \mbox{\boldmath $1$}
\end{array}
\right) .
\end{equation}
The span list for $\tilde H'$ is given by
\begin{equation}
\hat S=\{(2, 10], (5, 13], (8, 1], (11, 4], (14, 7]\} .
\end{equation}
We see that $\tilde H'$ is equivalent to $\tilde H_5^{tb}$. Thus $\tilde H'$ is a scalar check matrix corresponding to $G'$. Also, note that $\tilde H'$ is the TBGM associated with $\tilde H'(D)=(D+D^2, D^3, 1)$. We already know that $G'$ is the TBGM associated with $G'(D)$. Hence, by Proposition 5.3, a reciprocal dual encoder associated with $G'(D)$ is given by
\begin{equation}
\tilde H'(D)=(D+D^2, D^3, 1) .
\end{equation}
Observe that $\tilde H'(D)=(D+D^2, D^3, 1)$ has a factor $D$ in the first column and a factor $D^2$ in the second column. Then sweeping these factors out of the corresponding columns, the constraint length of $\tilde H'(D)$ is reduced to one. This fact implies that the constraint length of $G'(D)$ can also be reduced.
\par
In the following, we will show that reduction of $G'(D)$ is actually realized. For the purpose, a check matrix corresponding to $G'(D)$, i.e.,
\begin{equation}
H'(D)=(D^{-1}+D^{-2}, D^{-3}, 1)\times D^3=(D+D^2, 1, D^3)
\end{equation}
is used.
\par
Let $G(D)$ and $H(D)$ be a generator matrix and a corresponding check matrix for a convolutional code, respectively. In the following, this relation is denoted by $G(D) \Leftrightarrow H(D)$. It is shown~\cite{taji 112} that $G(D)$ and $H(D)$ can be reduced simultaneously, if reduction is possible, where the relation $\Leftrightarrow$ is retained in the whole reduction process. We apply the method to our case under consideration. 
\par
{\it Step 1:} For $G'(D)$, add the first row multiplied by $D$ to the second row. By Proposition 4.3, this is a TB-equivalent transformation. As a result, we have
\begin{displaymath}
G''(D)=\left(
\begin{array}{ccc}
1+D & D & 1 \\
D & D^2 & 1
\end{array}
\right)\Leftrightarrow H'(D)=(D+D^2, 1, D^3) .
\end{displaymath}
\par
{\it Step 2:} Divide the second column of $G''(D)$ by $D$, while divide the first and third columns of $H'(D)$ by $D$. Then we have
\begin{displaymath}
G'''(D)=\left(
\begin{array}{ccc}
1+D & 1 & 1 \\
D & D & 1
\end{array}
\right)\Leftrightarrow H'''(D)=(1+D, 1, D^2) .
\end{displaymath}
\par
{\it Step 3:} Multiply the third column of $G'''(D)$ by $D$, while divide the third column of $H'''(D)$ by $D$. Then we have
\begin{displaymath}
G^{(4)}(D)=\left(
\begin{array}{ccc}
1+D & 1 & D \\
D & D & D
\end{array}
\right)\Leftrightarrow H^{(4)}(D)=(1+D, 1, D) .
\end{displaymath}
\par
{\it Step 4:} Note that $G^{(4)}(D)=\left(
\begin{array}{ccc}
1+D & 1 & D \\
D & D & D
\end{array}
\right)$ is not basic~\cite{forn 70}. Using an invariant-factor decomposition~\cite{forn 70} of $G^{(4)}(D)$, an equivalent basic matrix
\begin{equation}
G^{(5)}(D)=\left(
\begin{array}{ccc}
1+D & 1 & D \\
1 & 1 & 1
\end{array}
\right)
\end{equation}
is obtained. Note that $G^{(4)}(D)$ and $G^{(5)}(D)$ are TB-equivalent (cf. Proposition 4.4).
\par
In the above reduction process for $G'(D)$, except for TB-equivalent transformations, the second column is divided by $D$, whereas the third column is multiplied by $D$. Accordingly, for each TB path, let us cyclically shift the second component of each branch to the left by one branch and cyclically shift the third component of each branch to the right by one branch. Then the modified TB paths are represented as a TB convolutional code defined by $G^{(5)}(D)$ (see Propositions 4.1 and 4.2). The trellis for $G^{(5)}(D)$ is shown in Fig.3. For example, take an information sequence
\begin{displaymath}
\mbox{\boldmath $u$}=01~~00~~01~~00~~00
\end{displaymath}
and the corresponding TB path
\begin{displaymath}
\mbox{\boldmath $w$}=\mbox{\boldmath $u$}G_5^{tb}=011~~001~~110~~001~~101 .
\end{displaymath}
By cyclically shifting the second component of each branch to the left by one branch, and by cyclically shifting the third component of each branch to the right by one branch, we have
\begin{displaymath}
\mbox{\boldmath $w$}_m=001~~011~~101~~000~~111 .
\end{displaymath}
We see that $\mbox{\boldmath $w$}_m$ is a TB path which starts and ends in state $(0)$ in Fig.3.
\begin{figure}[htb]
\begin{center}
\includegraphics[width=8.0cm,clip]{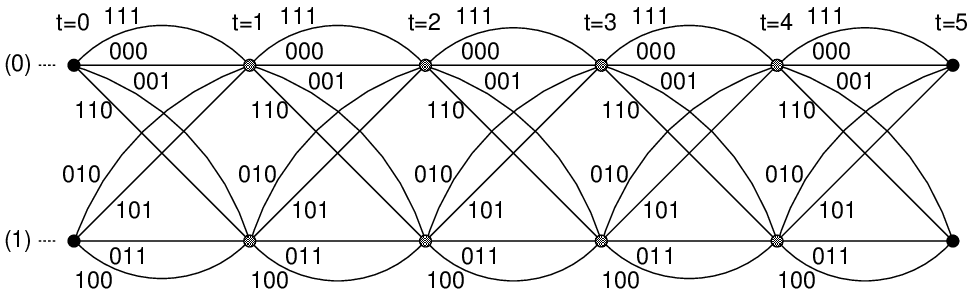}
\end{center}
\caption{The tail-biting convolutional code defined by $G^{(5)}(D)$~($N=5$).}
\label{Fig.3}
\end{figure}
\par
{\it Remark:} We remark that in the above argument, it is assumed that $G'$ is equivalent to $G_5^{tb}$ (i.e., the equivalence has been checked beforehand). In general, however, $k=k_0N$ is relatively large for high rate codes. Hence, it is preferable that the equivalence of $G'$ and $G_5^{tb}$ is derived without checking it beforehand. Actually, the equivalence of $G'$ and $G_5^{tb}$ is derived from the equivalence of $\tilde H'$ and $\tilde H_5^{tb}$ using the result of Gluesing-Luerssen and Weaver~\cite[Theorem IV.3]{glu 112} (see Appendix A).


\subsection{Relation Between Trellis Reduction and Section Length}
In the proposed trellis reduction method, the section length $N$ is an important parameter. Actually, the method is effective for TB convolutional codes of short to moderate section length. This is because the span lengths of characteristic generators increase as $N$ grows (see Section III-D). We have already shown that a TB trellis with generator matrix $G(D)=(1+D+D^2, 1+D^2)$ can be reduced for the case of $N=5$. Consider the same trellis. This time, however, $N$ is set to $6$. Then $G_6^{tb}$ is given by
\begin{eqnarray}
G_6^{tb} &=& \left(
\begin{array}{cccccc}
G_0 & G_1 & G_2 & 0 & 0 & 0  \\
0 & G_0 & G_1 & G_2 & 0 & 0  \\
0 & 0 & G_0 & G_1 & G_2 & 0 \\
0 & 0 & 0 & G_0 & G_1 & G_2 \\
G_2 & 0 & 0 & 0 & G_0 & G_1 \\
G_1 & G_2 & 0 & 0 & 0 & G_0
\end{array}
\right) \\
&=& \left(
\begin{array}{cccccccccccc}
\mbox{\boldmath $1$} & \mbox{\boldmath $1$} & \mbox{\boldmath $1$} & \mbox{\boldmath $0$} & \mbox{\boldmath $1$} & \mbox{\boldmath $1$} & 0 & 0 & 0 & 0 & 0 & 0 \\
0 & 0 & \mbox{\boldmath $1$} & \mbox{\boldmath $1$} & \mbox{\boldmath $1$} & \mbox{\boldmath $0$} & \mbox{\boldmath $1$} & \mbox{\boldmath $1$} & 0 & 0 & 0 & 0 \\
0 & 0 & 0 & 0 & \mbox{\boldmath $1$} & \mbox{\boldmath $1$} & \mbox{\boldmath $1$} & \mbox{\boldmath $0$} & \mbox{\boldmath $1$} & \mbox{\boldmath $1$} & 0 & 0 \\
0 & 0 & 0 & 0 & 0 & 0 & \mbox{\boldmath $1$} & \mbox{\boldmath $1$} & \mbox{\boldmath $1$} & \mbox{\boldmath $0$} & \mbox{\boldmath $1$} & \mbox{\boldmath $1$} \\
\mbox{\boldmath $1$} & \mbox{\boldmath $1$} & 0 & 0 & 0 & 0 & 0 & 0 & \mbox{\boldmath $1$} & \mbox{\boldmath $1$} & \mbox{\boldmath $1$} & \mbox{\boldmath $0$} \\
\mbox{\boldmath $1$} & \mbox{\boldmath $0$} & \mbox{\boldmath $1$} & \mbox{\boldmath $1$} & 0 & 0 & 0 & 0 & 0 & 0 & \mbox{\boldmath $1$} & \mbox{\boldmath $1$}
\end{array}
\right)
\begin{array}{l}
(0, 5] \\
(2, 7] \\
(4, 9] \\
(6, 11] \\
(8, 1] \\
(10, 3] .
\end{array}
\end{eqnarray}
Note that to each generator in $G_6^{tb}$, its span is assigned in the natural manner. Observe that the span lengths of these spans are the same, i.e., $6$. A characteristic matrix is computed as follows:
\begin{equation}
X=\left(
\begin{array}{cccccccccccc}
\mbox{\boldmath $1$} & \mbox{\boldmath $1$} & \mbox{\boldmath $1$} & \mbox{\boldmath $0$} & \mbox{\boldmath $1$} & \mbox{\boldmath $1$} & 0 & 0 & 0 & 0 & 0 & 0 \\
0 & \mbox{\boldmath $1$} & \mbox{\boldmath $0$} & \mbox{\boldmath $1$} & \mbox{\boldmath $1$} & \mbox{\boldmath $1$} & \mbox{\boldmath $1$} & \mbox{\boldmath $1$} & \mbox{\boldmath $1$} & 0 & 0 & 0 \\
0 & 0 & \mbox{\boldmath $1$} & \mbox{\boldmath $1$} & \mbox{\boldmath $1$} & \mbox{\boldmath $0$} & \mbox{\boldmath $1$} & \mbox{\boldmath $1$} & 0 & 0 & 0 & 0 \\
0 & 0 & 0 & \mbox{\boldmath $1$} & \mbox{\boldmath $0$} & \mbox{\boldmath $1$} & \mbox{\boldmath $1$} & \mbox{\boldmath $1$} & \mbox{\boldmath $1$} & \mbox{\boldmath $1$} & \mbox{\boldmath $1$} & 0 \\
0 & 0 & 0 & 0 &  \mbox{\boldmath $1$} & \mbox{\boldmath $1$} & \mbox{\boldmath $1$} & \mbox{\boldmath $0$} & \mbox{\boldmath $1$} & \mbox{\boldmath $1$} & 0 & 0 \\
\mbox{\boldmath $1$} & 0 & 0 & 0 & 0 & \mbox{\boldmath $1$} & \mbox{\boldmath $0$}& \mbox{\boldmath $1$} & \mbox{\boldmath $1$} & \mbox{\boldmath $1$} & \mbox{\boldmath $1$} & \mbox{\boldmath $1$} \\
0 & 0 & 0 & 0 & 0 & 0 & \mbox{\boldmath $1$} & \mbox{\boldmath $1$} & \mbox{\boldmath $1$} & \mbox{\boldmath $0$} & \mbox{\boldmath $1$} & \mbox{\boldmath $1$}\\
\mbox{\boldmath $1$} & \mbox{\boldmath $1$} & \mbox{\boldmath $1$} & 0 & 0 & 0 & 0 & \mbox{\boldmath $1$} & \mbox{\boldmath $0$}& \mbox{\boldmath $1$} & \mbox{\boldmath $1$} & \mbox{\boldmath $1$}\\
\mbox{\boldmath $1$} & \mbox{\boldmath $1$} & 0 & 0 & 0 & 0 & 0 & 0 & \mbox{\boldmath $1$} & \mbox{\boldmath $1$} & \mbox{\boldmath $1$} & \mbox{\boldmath $0$} \\
\mbox{\boldmath $1$} & \mbox{\boldmath $1$} & \mbox{\boldmath $1$} & \mbox{\boldmath $1$} & \mbox{\boldmath $1$} & 0 & 0 & 0 & 0 & \mbox{\boldmath $1$} & \mbox{\boldmath $0$} & \mbox{\boldmath $1$} \\
\mbox{\boldmath $1$} & \mbox{\boldmath $0$} & \mbox{\boldmath $1$} & \mbox{\boldmath $1$} & 0 & 0 & 0 & 0 & 0 & 0 & \mbox{\boldmath $1$} & \mbox{\boldmath $1$} \\
\mbox{\boldmath $0$} & \mbox{\boldmath $1$} & \mbox{\boldmath $1$} & \mbox{\boldmath $1$} & \mbox{\boldmath $1$} & \mbox{\boldmath $1$} & \mbox{\boldmath $1$} & 0 & 0 & 0 & 0 & \mbox{\boldmath $1$}
\end{array}
\right)
\begin{array}{l}
(0, 5] \\
(1, 8] \\
(2, 7] \\
(3, 10] \\
(4, 9] \\
(5, 0] \\
(6, 11] \\
(7, 2] \\
(8, 1] \\
(9, 4] \\
(10, 3] \\
(11, 6] .
\end{array}
\end{equation}
Thus the set of basic spans is given by
\begin{equation}
T_0=\left\{
\begin{array}{l}
(0, 5] \\
(1, 8] .
\end{array} \right.
\end{equation}
With respect to $G'$, there are two cases. When the first row of $X$ is used as a basic generator of $G'$, $G'$ is identical to $G_6^{tb}$. When the second row of $X$ is used as a basic generator of $G'$, the span lengths of rows of $G'$ are $8$ and are greater than $6$. These facts mean that in either case, trellis reduction is not realized using the proposed method. On the other hand, this example implies that the upper bound for $N$ can be estimated by comparing the span lengths of generators in $G'$ with those of generators in $G_N^{tb}$.
\par
Let $X$ be a characteristic matrix for a TB convolutional code of section length $N$. We already know that the associated span list $T$ consists of the set of basic spans
\begin{displaymath}
T_0=\left\{
\begin{array}{l}
(0, b_0] \\
(1, b_1] \\
\cdots \\
(n_0-1, b_{n_0-1}]
\end{array} \right.
\end{displaymath}
and $\rho_{in_0}(T_0)~(i=1, 2, \cdots, N-1)$. Also, the sum of span lengths of spans in $T_0$ is given by
\begin{displaymath}
\ell=n_0((n_0-k_0)N+1) .
\end{displaymath}
In the proposed method, $G'$ consists of $k$ generators in $X$. From a span viewpoint, this corresponds to choosing $k_0$ spans from $T_0$. Accordingly, the sum of span lengths of these $k_0$ spans, denoted by $\ell'$, is approximated by
\begin{eqnarray}
\ell' &\fallingdotseq& (k_0/n_0) \times n_0((n_0-k_0)N+1) \nonumber \\
&=& k_0((n_0-k_0)N+1) .
\end{eqnarray}
\par
On the other hand, consider $G_N^{tb}$, where to each generator, its span is assigned in the natural manner. Then the span list consists of the set of basic spans $\hat T_0$ and $\rho_{in_0}(\hat T_0)~(i=1, 2, \cdots, N-1)$. We evaluate the sum of span lengths of spans in $\hat T_0$, denoted by $\hat \ell$. Let $\nu_i$ be the degree of the $i$th row of $G(D)$. Here take notice of the first block of $k_0$ rows in $G_N^{tb}$, i.e., 
\begin{displaymath}
\left(G_0, G_1, \cdots, G_{L-1}, G_L, 0, \cdots, 0 \right) .
\end{displaymath}
The span length of the $i$th ($1 \leq i \leq k_0$) row is approximated by $n_0(\nu_i+1)$. Hence, we have
\begin{eqnarray}
\hat \ell &\fallingdotseq& n_0(\nu_1+1)+n_0(\nu_2+1)+ \cdots +n_0(\nu_{k_0}+1) \nonumber \\
&=& n_0(\nu_1+\nu_2+ \cdots +\nu_{k_0})+n_0k_0 \nonumber \\
&=& n_0(\nu+k_0) ,
\end{eqnarray}
where $\nu\stackrel{\triangle}{=}\nu_1+\nu_2+ \cdots +\nu_{k_0}$ is the constraint length of $G(D)$. Since trellis reduction is realized in the case where $G'$ consists of generators with short span length, we can take the inequality
\begin{displaymath}
\ell' \leq \hat \ell
\end{displaymath}
as a criterion for trellis reduction. That is, we can estimate the upper bound for $N$ using the inequality
\begin{equation}
(\sharp)~~k_0((n_0-k_0)N+1) \leq n_0(\nu+k_0) .
\end{equation}
\par
For several concrete cases, we will show the condition $(\sharp)$.
\begin{itemize}
\item[(1)] $R=1/2$:
\begin{displaymath}
\left\{
\begin{array}{c}
\ell' \fallingdotseq N+1 \\
\hat \ell \fallingdotseq 2\nu+2
\end{array} \right.
\end{displaymath}
\begin{displaymath}
(\sharp)~N \leq 2\nu+1 .
\end{displaymath}
\item[(2)] $R=1/3$:
\begin{displaymath}
\left\{
\begin{array}{c}
\ell' \fallingdotseq 2N+1 \\
\hat \ell \fallingdotseq 3\nu+3
\end{array} \right.
\end{displaymath}
\begin{displaymath}
(\sharp)~N \leq \lfloor(3/2)\nu \rfloor+1 .
\end{displaymath}
\item[(3)] $R=2/3$:
\begin{displaymath}
\left\{
\begin{array}{c}
\ell' \fallingdotseq 2N+2 \\
\hat \ell \fallingdotseq 3\nu+6
\end{array} \right.
\end{displaymath}
\begin{displaymath}
(\sharp)~N \leq \lfloor(3/2)\nu \rfloor+2 .
\end{displaymath}
\end{itemize}
\par
We observe that the TB convolutional codes presented in Section V-B all satisfy the condition $(\sharp)$. Also, the rate $R=2/3$ TB convolutional code discussed in the previous section satisfies the condition $(\sharp)$.

\section{Conclusion}
In this paper, we have derived several basic properties of a characteristic matrix for a TB convolutional code. We have shown that the characteristic span list consists of some basic spans and their right cyclic shifts. Using the derived results, we have shown that a trellis associated with a given TB convolutional code can be reduced in some cases. As candidates for trellis reduction, we have taken the generator matrices from the tables in~\cite[Chapter 8]{joha 99} in principle. For example, the rate $R=1/2$ encoders in Section V-B were chosen from~\cite[TABLE 8.1]{joha 99}. On the other hand, good TB convolutional encoders have been obtained~\cite{joha 99,sta 99}. Here, for a given rate $R=k_0/n_0$, the optimal encoder of memory length $L$ produces the largest minimum distance $d$ for each section length $N$. We have applied the proposed reduction method to some of such encoders (see~\cite[TABLE 8.19]{joha 99}). As a result, for example, we have obtained $G=(6, 7)$ ($\nu=2,~N=5$) from $G=(50, 64)$ ($\nu=3,~N=5$), where the octal notation for generator matrices is used. Similarly, we have obtained $G=(54, 60)$ ($\nu=3,~N=6$) from $G=(46, 60)$ ($\nu=4,~N=6$). Note that both $G=(6, 7)$ ($\nu=2,~N=5$) and $G=(54, 60)$ ($\nu=3,~N=6$) are listed in the same table.
\par
Finally, we remark that the proposed trellis reduction method depends on the choice of a characteristic matrix for a given convolutional code. Though the number of characteristic matrices to be examined is rather restricted (cf. Proposition 5.1), the method is not fully constructive. Also, a detailed condition that trellis reduction is realized has to be clarified.


%

\appendices
\section{Proof of the equivalence of $G'$ and $G_5^{tb}$}
We first prove the following.
\begin{pro}
Let $G(D)$ be as in Section III. Consider a TB convolutional code $C$ generated by $G_N^{tb}$ and the corresponding dual code $C^{\perp}$. Let $X$ be a characteristic matrix for $C$ with span list $T$. Also, let $Y$ be a characteristic matrix for $C^{\perp}$ with span list $\hat T$. Let $\tilde X$ and $\tilde Y$ be submatrices of $X$ and $Y$, respectively, where $\tilde X$ consists of $k$ rows in $X$, whereas $\tilde Y$ consists of $(n-k)$ rows in $Y$. Denote by $S$ and $\hat S$ the span lists for $\tilde X$ and $\tilde Y$, respectively. Here assume the following:
\begin{itemize}
\item[i)] $\tilde X$ consists of $k_0$ rows and their right cyclic shifts by integer multiple of $n_0$.
\item[ii)] Each span in $S$ does not include any spans in $T$ except itself.
\end{itemize}
\begin{itemize}
\item[iii)] The rows of $\tilde Y$ are linearly independent and thus generate $C^{\perp}$.
\item[iv)] $\tilde Y$ consists of $(n_0-k_0)$ rows and their right cyclic shifts by integer multiple of $n_0$.
\item[v)] Each span in $\hat S$ does not include any spans in $\hat T$ except itself.
\end{itemize}
\begin{itemize}
\item[vi)] $S$ and $\hat S$ satisfy $(b, a]\in \hat S \leftrightarrow (a, b] \notin S$ (that is, $\hat S$ consists of the spans in $\hat T$ whose reverse is not in $S$).
\end{itemize}
Then the rows of $\tilde X$ are linearly independent, thus generate $C$, i.e., $\tilde X$ is equivalent to $G_N^{tb}$.
\end{pro}
\par
{\it Remark:} When $\tilde X$ consists of generators in $X$ with short span length, it is probable that the condition ii) holds. Similarly, when $\tilde Y$ consists of generators in $Y$ with short span length, it is probable that the condition v) holds.
\begin{IEEEproof}
From ii), it follows that $\tilde X$ is common to all the characteristic matrices for $C$. Similarly, from v), it follows that $\tilde Y$ is common to all the characteristic matrices for $C^{\perp}$.  Also, by vi), $(\tilde X, \tilde Y)$ is a {\it dual selection} of $(X, Y)$~\cite[Definition IV.2]{glu 112}. As a result~\cite[Theorem IV.3]{glu 112}, we have
\begin{itemize}
\item[1)]$\mbox{rank}~\tilde X=k \leftrightarrow \mbox{rank}~\tilde Y=n-k$
\item[2)]Let $\mbox{rank}~\tilde X=k$. Then the KV trellises~\cite{glu 112} based on $(\tilde X, S)$ and $(\tilde Y, \hat S)$ are dual to each other.
\end{itemize}
By iii), $\mbox{rank}~\tilde Y=n-k$. Hence, by 1), $\mbox{rank}~\tilde X=k$.
\end{IEEEproof}
\par
Let us go back to Example 4. In this example, the code $C$ generated by $G_5^{tb}$ and the dual code $C^{\perp}$ generated by $\tilde H_5^{tb}$ are considered. $X$ is a characteristic matrix for $C$, whereas $Y$ is a characteristic matrix for $C^{\perp}$. Note that neither $X$ nor $Y$ are unique. These are observed from the relation of inclusion in the associated span lists $T$ and $\hat T$, where
\begin{eqnarray}
T &=& \{(0, 4], (1, 8], (2, 6], (3, 7], (4, 11], (5, 9], (6, 10], (7, 14], \nonumber \\
&& (8, 12], (9, 13], (10, 2], (11, 0], (12, 1], (13, 5], (14, 3]\} \nonumber \\
\hat T &=& \{(0, 11], (1, 12], (2, 10], (3, 14], (4, 0], (5, 13], (6, 2], (7, 3], \nonumber \\
&& (8, 1], (9, 5], (10, 6], (11, 4], (12, 8], (13, 9], (14, 7]\} . \nonumber
\end{eqnarray}
\par
Next, take notice of the matrices $G'$ and $\tilde H'$, which are submatrices of $X$ and $Y$, respectively. The corresponding span lists are given by
\begin{eqnarray}
S &=& \{(0, 4], (2, 6], (3, 7], (5, 9], (6, 10], \nonumber \\
&& (8, 12], (9, 13], (11, 0], (12, 1], (14, 3]\} \nonumber \\
\hat S &=& \{(2, 10], (5, 13], (8, 1], (11, 4], (14, 7]\} . \nonumber
\end{eqnarray}
Here note the following:
\begin{itemize}
\item Each span in $S$ does not include any spans in $T$ except itself.
\item Each span in $\hat S$ does not include any spans in $\hat T$ except itself.
\end{itemize}
Moreover, $\hat S$ consists of the spans in $\hat T$ whose reverse is not in $S$. Actually, by reversing the spans in $\hat S$, we have
\begin{displaymath}
(10, 2], (13, 5], (1, 8], (4, 11], (7, 14].
\end{displaymath}
We see that these spans are not in $S$.
\par
All these facts show that the conditions in Proposition A.1 are satisfied, when $\tilde X$ and $\tilde Y$ are replaced by $G'$ and $\tilde H'$, respectively. Hence, the equivalence of $G'$ and $G_5^{tb}$ is derived.
\par
{\it Remark:}~\cite[Theorem IV.3]{glu 112} holds only for a pair $(X, Y)$, where $X$ is a characteristic matrix for $C$ and $Y$ is the corresponding dual one for $C^{\perp}$ (see~\cite{glu 112}). On the other hand, the pair $(X, Y)$ computed above may not be in the duality relation. However, $G'$ is common to all the characteristic matrices for $C$, and $\tilde H'$ is common to all the characteristic matrices for $C^{\perp}$ as well. Hence, the theorem can be applied to our case.



\section*{Acknowledgment}

The author would like to thank Prof. Heide Gluesing-Luerssen for valuable comments on the duality of Koetter-Vardy (KV) trellises.

\ifCLASSOPTIONcaptionsoff
  \newpage
\fi

\begin{IEEEbiographynophoto}{Masato Tajima}
(M'86--SM'13) was born in Toyama, Japan, on August 13, 1949. He received the B.E., M.E., and Dr. of Eng. degrees all in electrical engineering from Waseda University, Tokyo, Japan, in 1972, 1974, and 1979, respectively. He joined the Electronics Equipment Laboratory of Toshiba R$\&$D Center in 1979, where he engaged in research and development of channel coding techniques with applications to satellite communication systems. From 1993 to 2006, he was with the Department of Intellectual Information Systems Engineering, Toyama University, first as an Associate Professor, next as a Professor. From 2006 to 2015, he was with the Graduate School of Science and Engineering, University of Toyama, as a Professor. He is currently a Professor Emeritus at University of Toyama. His research interests are in coding theory and its applications.
\end{IEEEbiographynophoto}






\end{document}